\documentclass[twocolumn,showpacs,preprintnumbers,amsmath,amssymb]{revtex4}

\usepackage{graphicx}
\usepackage{dcolumn}
\usepackage{bm}
\newcommand{\be}{\begin{equation}}
\newcommand{\ee}{\end{equation}}
\let\del=\nabla
\newcommand{\bea}{\begin{eqnarray}}
\newcommand{\eea}{\end{eqnarray}}

\def\apj{ApJ}

\def\apjs{{ApJ\ Suppl.}}

\def\mnras{{MNRAS}}

\def\prd{{Phys. Rev. D}}
\def\physrep{{Phys.~Rep.}}   

\begin{document}


\title{Can the WMAP Haze really be a signature of annihilating neutralino dark matter?}

\author{Daniel T. Cumberbatch}
 \affiliation{CERCA, Physics Department, Case Western Reserve University, 10900 Euclid Avenue, Cleveland, OH 44106-7079, USA.}
 \email{dtc21@case.edu}
\author{Joe Zuntz}%
 \email{jaz@astro.ox.ac.uk}
 \affiliation{
Department of Astrophysics, University of Oxford, Denys Wilkinson Building, Keble Road, Oxford, OX1 3RH.}%
\author{Hans Kristian Kamfjord Eriksen}%
 \email{h.k.k.eriksen@astro.uio.no}
 \affiliation{Institute of Theoretical Astrophysics, University of Oslo, P.O. Box 1029 Blindern, N-0315 Oslo, Norway.
}%
\author{Joe Silk}%
 \email{silk@astro.ox.ac.uk}
  \affiliation{
Department of Astrophysics, University of Oxford, Denys Wilkinson Building, Keble Road, Oxford, OX1 3RH.}%

\date{\today}

\begin{abstract}
Observations by the Wilkinson Microwave Anisotropy Probe (WMAP) satellite have identified an excess of microwave emission from the centre of the Milky Way. It has been suggested that this {\it WMAP haze} emission could potentially be synchrotron emission from relativistic electrons and positrons produced in the annihilations of one (or more) species of dark matter particles. In this paper we re-calculate the intensity and morphology of the WMAP haze using a multi-linear regression involving full-sky templates of the dominant forms of galactic foreground emission, using two different CMB sky signal estimators. The first estimator is a posterior mean CMB map, marginalized over a general foreground model using a Gibbs sampling technique, and the other is the ILC map produced by the WMAP team. Earlier analyses of the WMAP haze used the ILC map, which is more contaminated by galactic foregrounds than the Gibbs map. In either case, we re-confirm earlier results that a statistically significant residual emission remains after foreground subtraction that is concentrated around the galactic centre. However, we find that the significance of this emission can be significantly reduced by allowing for a subtle spatial variation in the frequency dependence of soft synchrotron emission in the inner and outer parts of the galaxy. We also re-investigate the prospect of a neutralino dark matter interpretation of the origin of the haze, and find that significant boosting in the dark matter annihilation rate is required, relative to that obtained with a smooth galactic dark matter distribution, in order to reproduce the inferred residual emission, contrary to that deduced in several recent studies.
\end{abstract}

\maketitle

\section{Introduction}
As the statistical error bars on Cosmic Microwave Background (CMB) measurements shrink, the importance of correct foreground subtraction grows.  The separation of the various forms of Galactic foreground emission from the primordial CMB signal is both critical to correct estimation of the CMB statistics and very difficult.  The application of foreground subtraction to the all-sky microwave measurements from the Wilkinson Microwave Anisotropy Probe (WMAP) has been extensively studied \cite{fg_subtraction}.  

In WMAP's frequency range $23<\nu<94\,$GHz the dominant sources of foreground signal come from emission in the interstellar medium: free-free (thermal bremsstrahlung), dust and synchrotron emission.  A correct separation of these sources, which are the dominant signal over large parts of the sky, allows both sensible use of the CMB data for cosmological parameter estimation and for an improved understanding of the emission mechanisms in the ISM.  The use of prior information about the likely spatial and frequency dependence of the sources makes this separation possible with current data.

During the study of Galactic foreground emission evidence has emerged for an additional, possibly exotic, emission mechanism responsible for an excess of microwave radiation emerging from the Galactic centre (GC) and distributed approximately symmetrically around it \cite{Finkbeiner2005:ch3}. This currently unexplained excess emission has been referred to as the WMAP haze, and its origin has been the subject of much debate. 

The WMAP Haze was originally suggested to be an excess in free-free emission, originating from hot ionised gas with temperature $T\sim10^5~$K, owing to its alleged hard spectrum \cite{Finkbeiner2004:ch3}.  However, such gas is thermally unstable, and with an insufficient abundance of gas at hotter or cooler temperatures, evident from constraints on the intensity of recombination lines/x-rays, such an explanation of the signal seems unlikely \cite{Finkbeiner2004:ch3, Spitzer1978:ch3}.
The hard spectrum has also led to speculation 
(see e.g.~\cite{HFD2007:ch3, Cholis_xdm, Caceres08, Cholis08}) that the origin of the haze is more likely to be hard synchrotron foreground emission resulting from a source distinct from that responsible for the softer synchrotron emission commonly associated with Galactic foregrounds. An astrophysical origin of such a hard synchrotron spectrum, such as that produced from a distribution of ultra-relativistic electrons, possibly resulting from supernovae explosions, with a flux large enough to account for the haze, seems difficult to realise \cite{HFD2007:ch3}.

Here, we re-evaluate the recent proposal that the WMAP haze may originate from synchrotron emission produced by a distribution of high-energy positrons and electrons resulting from the self-annihilation of supersymmetric neutralino dark matter. 
What differentiates our study from those past \cite{Finkbeiner2005:ch3, Finkbeiner2007a:ch3, HFD2007:ch3, Caceres08, Cholis08} is that in addition to using a standard set of templates, designed to model Galactic foreground emission, in a template-based, multi-linear fit to the CMB-subtracted WMAP data, we use a by-product of the {\it Gibbs sampling} method of CMB parameter estimation as our CMB contribution to the data \cite{Gibbs:ch3}, and compare our results with those calculated using the internal linear combination method (ILC) utilised in previous work \cite{Finkbeiner2005:ch3}. Gibbs sampling is a Monte-Carlo Markov Chain method that generates posterior samples of the signal map, power spectrum, and foreground components. The map from this method is the output of a well-understood statistical process that is known to produce good results, and may be one of the cleanest CMB temperature maps currently available.   

There have been many other proposals for the origin of the WMAP haze involving dark matter candidates other than supersymmetric neutralino dark matter. These include, exciting dark matter (XDM) particles \cite{Cholis_xdm} which annihilate to produce a hard spectrum of $e^{\pm}$ centred around the GC. Also, compact composite object (CCO) dark matter \cite{Forbes}, which are macroscopic assemblies of quarks in a colour superconducting state whose finite electric charge is neutralised by an atmosphere of $e^{\pm}$ bound to each CCO. It is then proposed that these $e^{\pm}$ annihilate with matter in the ISM to produce radiation which contributes to a number of unexplained astrophysical observations, including the WMAP Haze and the 511\,keV excess from the GC (however the prospects for a CCO explanation of the 511\,keV signal are considered by some to be unlikely \cite{Cumberbatch_CCO} ). Also, Sommerfeld enhanced dark matter has been proposed to explain the WMAP haze \cite{Lattanzi}, where the annihilation rate of TeV-mass particles are significantly boosted in low velocity environments, owing to an additional Yukawa-like attractive force between colliding particles.


\section{WMAP data and templates for Galactic foregrounds}
\label{sec:templates}
The three forms of Galactic foreground emission that dominate in the range of frequencies observed by WMAP are: free-free (or thermal bremsstrahlung) emission from hot ionised gas; dust emission, in two forms: firstly from {\it thermal  emissions} from dust particles interacting with the ambient radiation field, and secondly from {\it spinning dust} particles excited by collisions with ions traversing the ISM \cite{footnote1}, and synchrotron emission originating from electrons which have been shock-accelerated to GeV--energies by supernova explosions.
Since microwave radiation is not significantly attenuated by the particles involved in these emission mechanisms, the all-sky maps produced from the CMB-subtracted WMAP data should theoretically be reproducible from a linear combination of templates modelling the various foreground components, constructed from independent data maps of the three dominant components of foreground emission. But firstly, we briefly discuss the nature of the CMB-subtracted WMAP data that is to be used in our anlaysis.

\subsection{CMB-subtracted WMAP data}
\label{subsec:WMAPdata}
We utilise the 3yr WMAP data maps \cite{footnote2} provided in the HEALPix \cite{footnote3} pixelization scheme. We construct temperature maps at a given frequency using a weighted  
average of all the WMAP differencing array (DA) maps at that  
frequency.  The weighting, which is different at each pixel $p$, is  
given by $1/\sigma^2_{pi}$, where $\sigma_{pi} = \sigma_i/\sqrt{n_p} $  
is the error in the pixel for the given DA,  where $\sigma_i$ is the  
characteristic noise of the DA and $n_p$ is the number of times the  
pixel is measured.  Such data has been released by WMAP \cite{footnote4}.
Our CMB map \cite{footnote5} is the output of the Gibbs sampling method described, implemented and
applied to the 3-year WMAP data in \cite{Gibbs:ch3}. This CMB map was
produced by marginalizing the joint foreground-CMB posterior over a
free low-frequency foreground amplitude and spectral index at each
pixel, as well as a single amplitude relative to the dust template
described below \cite{Finkbeiner1999:ch3}. These additional degrees of
freedom allows for a much more accurate CMB reconstruction than that
provided by the WMAP ILC map. However, all results presented in this
paper are computed for both the Gibbs and the ILC maps, to assess the
impact of residual foregrounds on our results. In general, we find
only small differences between the two data sets, and no main
conclusions depend significantly on the CMB map choice.

We used the point source mask recently updated and distributed with the latest 5-year data release by WMAP \cite{Gold2008:ch3} to mask areas of the sky which may contaminate our fit. 


\subsection{Free-free template}
\label{subsec:fftemp:ch3}
Free-free (or thermal bremsstrahlung) emission originates from coulomb interactions between free electrons and hot interstellar gas. Since free-free emission is proportional to the emission measure EM$=\int n_e^2{\rm d}l$, where $n_e$ is the spatially-dependent number density of interstellar electrons,  
along a given line of sight, maps of H$\alpha$ recombination line emission (which is also proportional to the emission measure) can be used to approximately trace the morphology of free-free emission in the Galaxy. In fact, H$\alpha$ maps at intermediate and high latitudes are the only effective free-free templates available.
In this study we use the all-sky H$\alpha$ template described in \cite{Finkbeiner2003:ch3} which is a composite of data obtained from the WHAM Fabry-Perot survey of the northern sky \cite{Haffner2003:ch3} (which provides a good distinction between the Galactic emission and the geocoronal H$\alpha$ emission, which is caused by solar Ly-$\beta$ excitation of HI in the exosphere) the SHASSA filter survey of the southern sky \cite{Gaustad2001:ch3} and the variable resolution VTSS filter survey of the northern hemisphere \cite{Dennison1998:ch3}. 
\begin{figure}[h]
\vspace{0.5cm}
\begin{center}
\includegraphics[scale=0.25, bb= 100 100 510 720, angle=90]{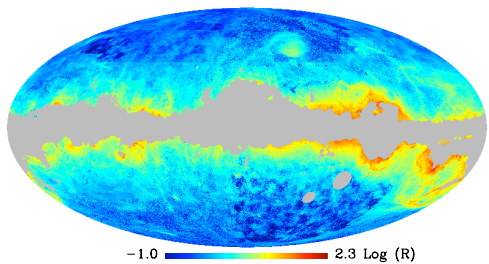}
\caption{The Free-free template map, masked using the WMAP five-year point source mask described in \S\,\ref{subsec:WMAPdata}, and presented in units of log$[{\rm Rayleighs}(R)]$, where $1R=10^6/4\pi$ photons cm$^{-2}$~s$^{-1}$~sr$^{-1}$.}
\label{fig:Ha_log}
\end{center}
\end{figure}
When using the H$\alpha$ map as a template for free-free emission, it is necessary to correct for extinction by dust (both in front of and mixed with the warm gas generating the H$\alpha$ flux). We correct our template for dust extinction by following the treatment in \cite{Finkbeiner2003:ch3, Finkbeiner2007a:ch3}: we assume that the warm gas is {\it uniformly} mixed with dust in a cloud possessing optical depth $\tau$, the observed spectral intensity $I_{\nu, {\rm obs.}}$ is then related to the emitted intensity $I_{\nu, {\rm em.}}$ by
\begin{equation}
I_{\nu, {\rm obs.}}=\frac{I_{\nu, {\rm em.}}}{\tau}\left(1-e^{-\tau}\right),
\end{equation}
where $\tau$ is computed by multiplying the all-sky 100\,$\mu$m dust map (here, in units of mK antenna temperature) produced by Schlegel, Finkbeiner and Davis (SFD henceforth) \cite{Schlegel1998:ch3}, in units of $E(B-V)$ magnitudes reddening, by a factor 2.65/1.086 \cite{Finkbeiner2003:ch3}. We then further reduce the emitted flux by masking out all regions where the dust extinction $A({\rm H}\alpha)\equiv2.65E(B-V)>1$. 

The conversion of dust-corrected H$\alpha$ intensities to EM and then to free-free emission intensities is well understood. The corresponding free-free brightness temperature $T_b$ is then related to the electron temperature and EM by
\be
T_b\propto T_e^{-0.35}\nu^{-2.15}\times{\rm EM},
\ee
where over the range of frequencies $\nu$ observed by WMAP, $T_e$ is approximately constant ($\simeq8000$K)\cite{Spitzer1978:ch3, DDD2003:ch3}.

The resulting template map is displayed in fig.~\ref{fig:Ha_log}


\subsection{Dust template}
\label{subsec:dusttemp:ch3}
Thermal dust emission is produced by microscopic interstellar dust grains vibrating in thermal equilibrium with the ambient radiation field. Here, we use the template map of dust emission by Finkbeiner, Davis and Schlegel (FDS henceforth), evaluated at 94\,GHz \cite{Finkbeiner1999:ch3}. It has also been proposed in \cite{Finkbeiner2007b:ch3} that such a map should closely trace the electric dipole emission expected to be produced by the smallest dust grains, which have a finite dipole moment, excited into rotational modes mainly by collisions with surrounding ions \cite{Draine:ch3} \cite{footnote6}.
\begin{figure}[h]
\vspace{0.5cm}
\begin{center}
\includegraphics[scale=0.25, bb= 100 100 510 720, angle=90]{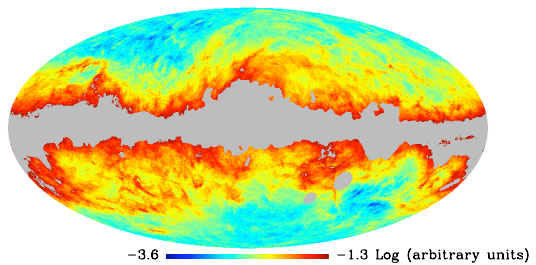}
\caption{The dust template map, masked using the WMAP five-year point source mask, and presented using arbitrary units.}
\label{fig:dust_log}
\end{center}
\end{figure}
The dust template map is displayed in fig.~\ref{fig:dust_log}.

\subsection{Synchrotron template}
\label{subsec:synctemp:ch3}
Galactic synchrotron emission is mainly produced by electrons in the vicinity of supernovae explosions which are shock-accelerated to relativistic (at least MeV) energies and then subsequently lose energy, mainly as inverse Compton radiation from scattering with CMB photons and starlight, and synchrotron radiation from their interaction with the Galactic magnetic field. This emission is best measured at low frequencies (i.e.~$<1$GHz ) where contamination from other sources, mainly free-free emission, is its lowest. A full-sky map of this emission was measured at 408\,MHz by Haslam $\it et~al.$ \cite{Haslam1982:ch3}, which we use here as our template of Galactic synchrotron foreground emission. 
\begin{figure}[h]
\vspace{0.5cm}
\begin{center}
\includegraphics[scale=0.25, bb= 100 100 510 720, angle=90]{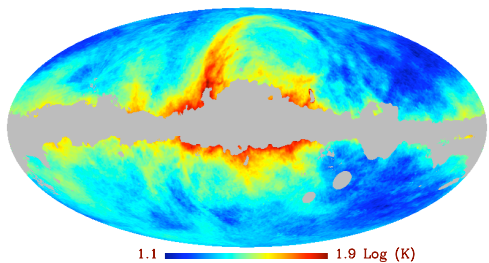}
\caption{The synchrotron template map, masked using the WMAP five-year point source mask, and presented in units log$[ T(K) ]$, where $T$ is antenna temperature.}
\label{fig:sync_log}
\end{center}
\end{figure}
This synchrotron template is displayed in fig.~\ref{fig:sync_log}.

The spectral dependence of the synchrotron brightness temperature, $T_b\propto\nu^{\beta}$, is expected to be highly spatially dependent, especially (hardening) near to recent  supernova activity. Past analyses of low frequency maps of Galactic synchrotron radiation have deduced wide-ranging values of the index $\beta$ consistent with the available data, with a value in the vicinity of -3 \cite{syncindex:ch3, Finkbeiner2007a:ch3}. 


\subsection{Template fitting procedure}
We perform a fit of the CMB-subtracted WMAP data described in \S\,\ref{subsec:WMAPdata} using maps in all five WMAP frequency bands (K, Ka, Q, V and W, corresponding to frequencies of 22.8, 33.0, 40.7, 60.8 and 93.5~GHZ respectively) by using a template-based multi-linear regression. We follow a treatment identical to that described in \cite{Finkbeiner2007a:ch3} and find the solution vector $\mathbf{a}$ for the template coefficients that minimizes the $\chi^2$ of the linear equation 
\be
P \mathbf{a} = \mathbf{w},
\ee
where $P$ describes the templates, $\mathbf{w}$ the data and $\mathbf{a}$ the coefficients of the templates in each frequency band.

We work with maps all downgraded to resolution $N_{\rm side}=64$ and smoothed using a beam of FWHM=3$^{\circ}$.  We extend the WMAP 5 year mask to cover the region where edge effects in the smoothing come into play.  We also manually remove point sources that appear as strong residuals during the analysis.

\section{3-template fit}
\label{sec:3temp}
In this section we demonstrate our template-based multi-linear fitting procedure by performing a basic 
linear regression involving the full-sky free-free, dust and synchrotron templates described and displayed in the previous section. 

We characterise the quality of our fits by using the statistic $\chi^2_{\rm red.}={\chi^2}/{\nu}$,
where $\chi^2$ is the Euclidean norm of the residual vector $\mathbf{r}=\mathbf{Pa}-\mathbf{w}$, divided by the WMAP 1$\sigma$ (noise) error maps $\mathbf{\sigma}$
\be
\chi^2=\left\|\frac{P}{\sigma}{\rm{\bf a}}-\frac{\rm{\bf w}}{\sigma}\right\|,
\label{eq:chisq}
\ee
and $\nu=N_p-N_a$, where $N_p$ is the number of unmasked pixels in each map and $N_a$ is the number of elements of {\bf a}.

\begin{figure}[h]
\vspace{0.5cm}
\begin{center}
\includegraphics[scale=0.17, bb= 100 200 510 700, angle=90]{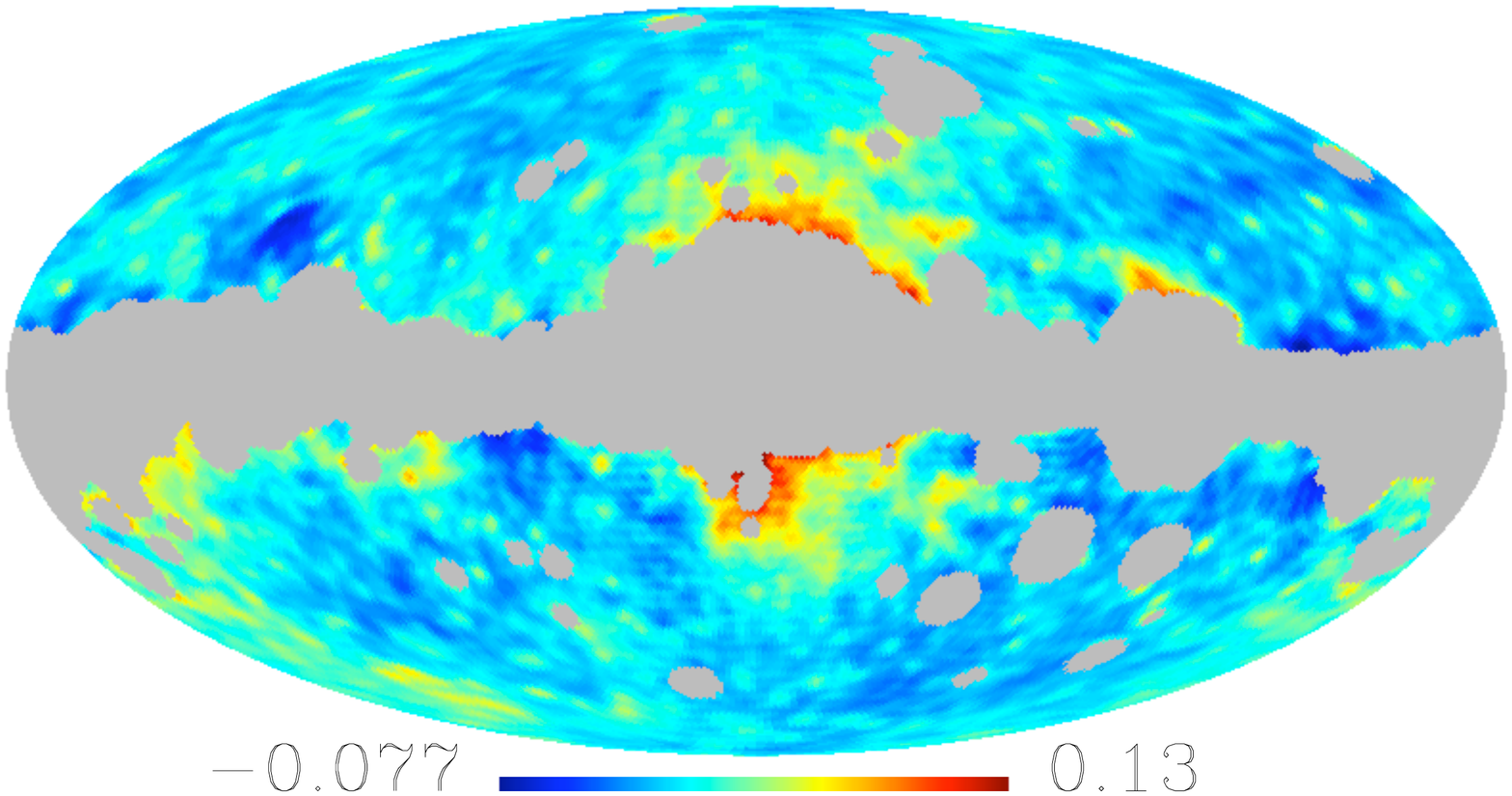}
\hspace{0.1cm}
\includegraphics[scale=0.17, bb= 100 150 510 900, angle=90]{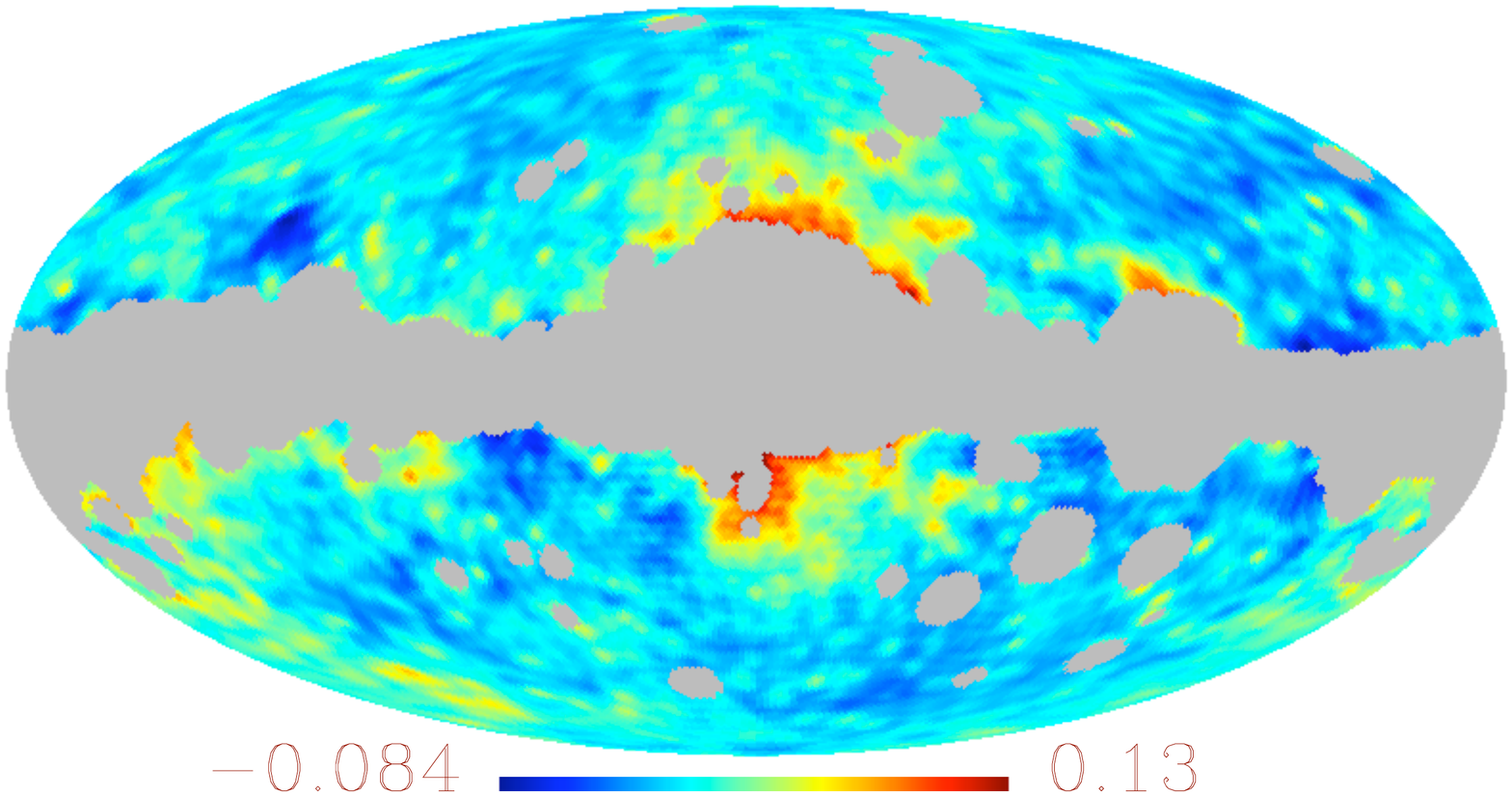}
\includegraphics[scale=0.17, bb= 100 200 510 700, angle=90]{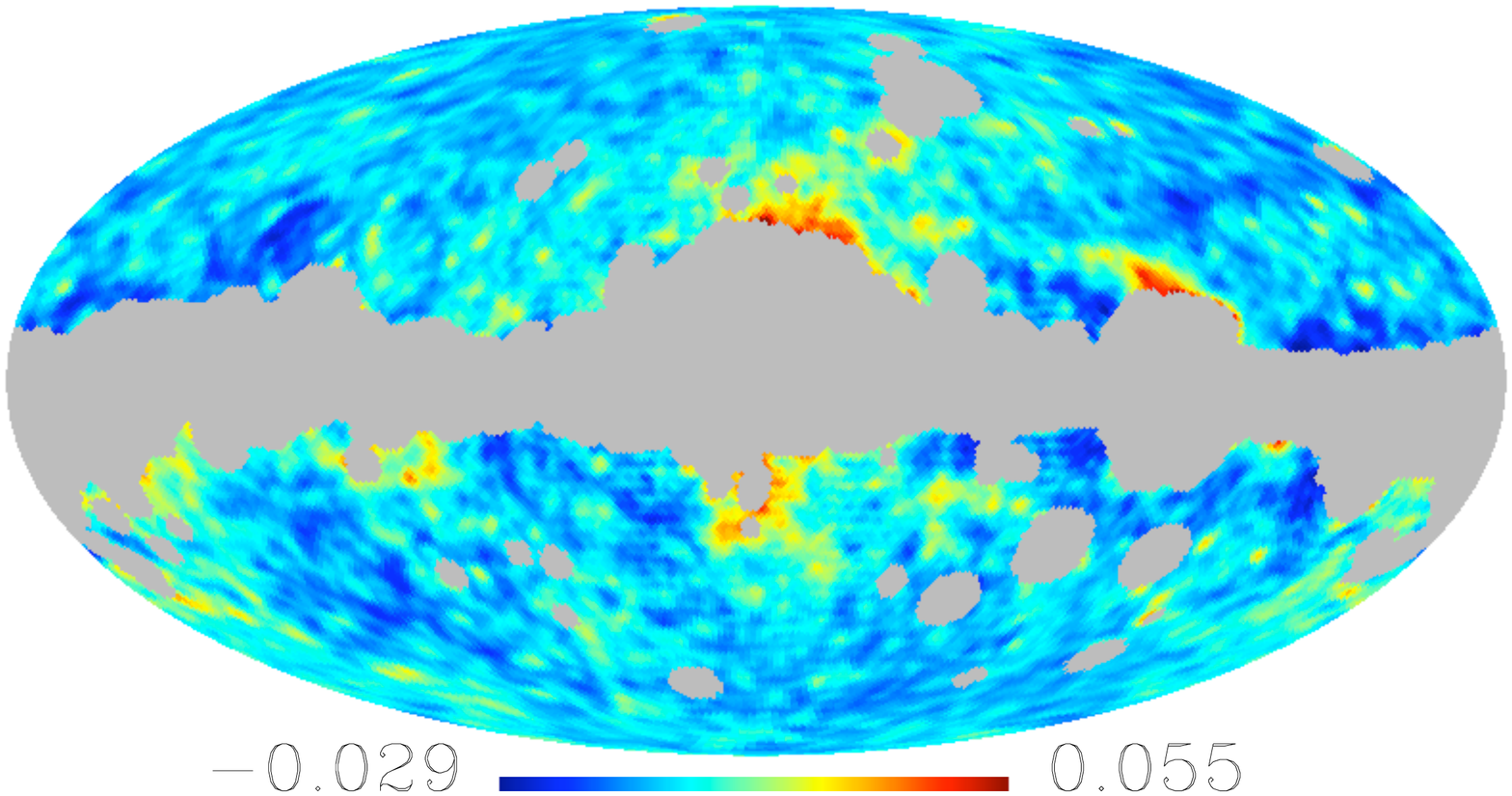}
\hspace{0.1cm}
\includegraphics[scale=0.17, bb= 100 150 510 900, angle=90]{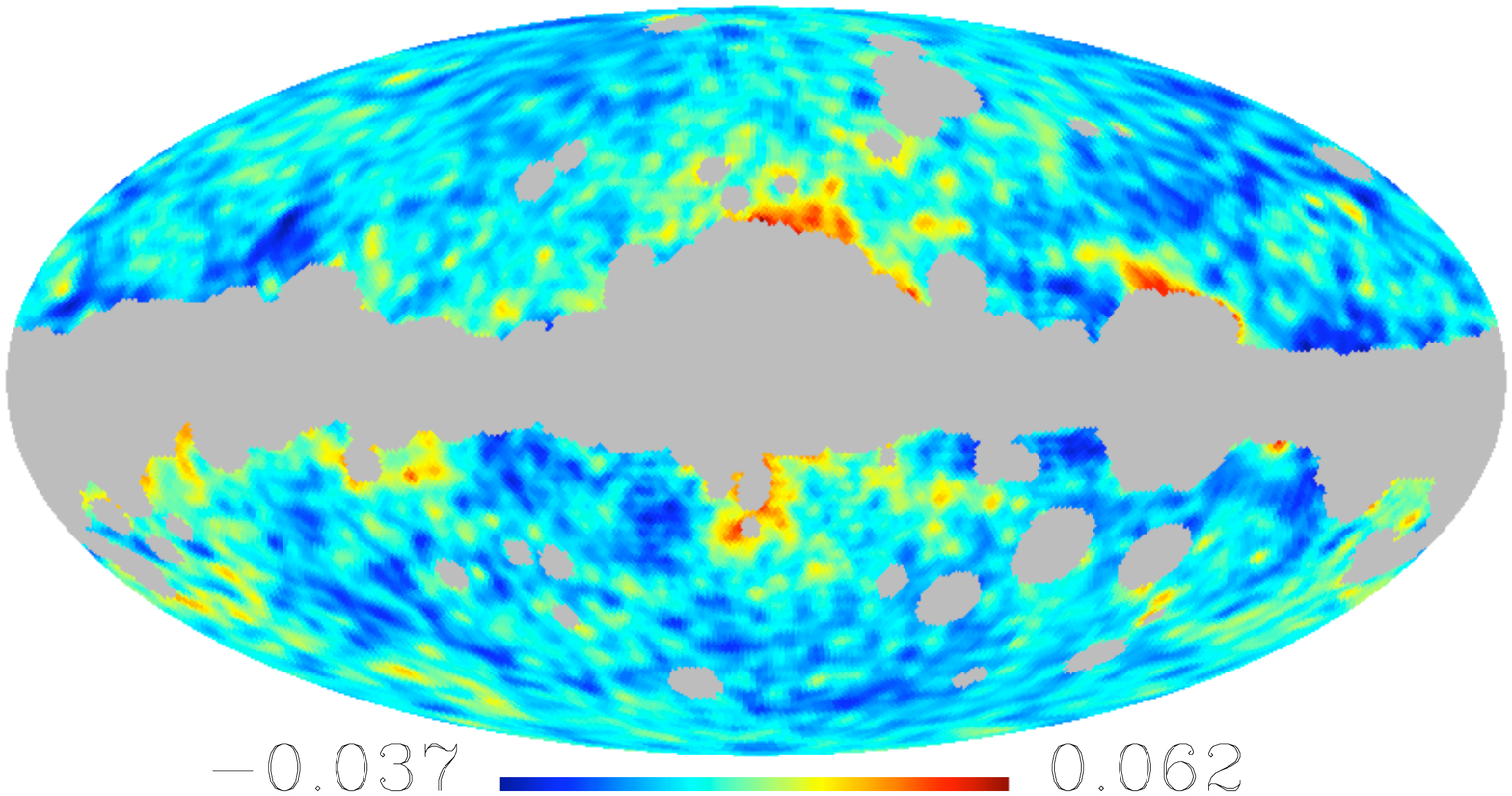}
\includegraphics[scale=0.17, bb= 100 200 510 700, angle=90]{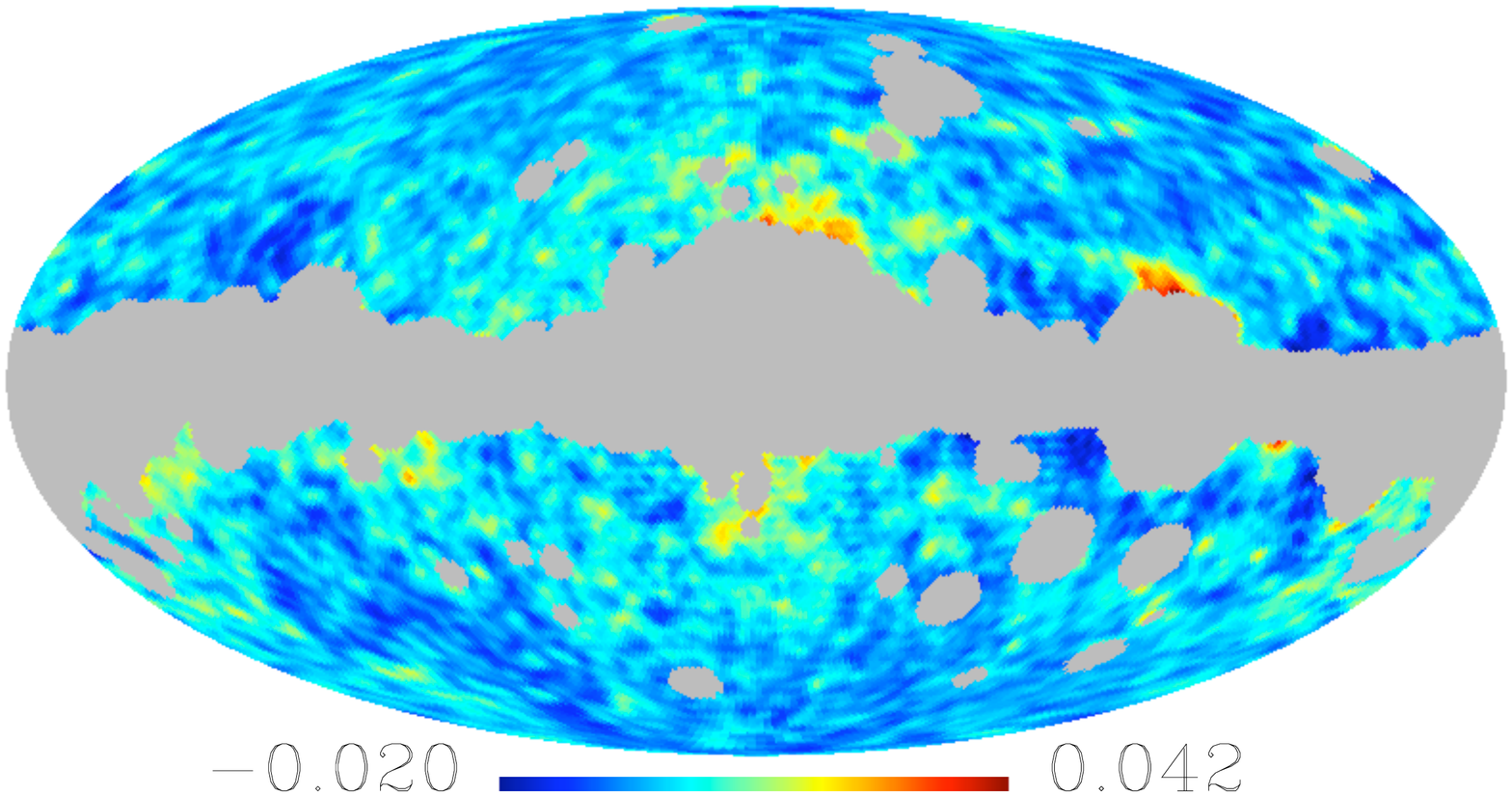}
\hspace{0.1cm}
\includegraphics[scale=0.17, bb= 100 150 510 900, angle=90]{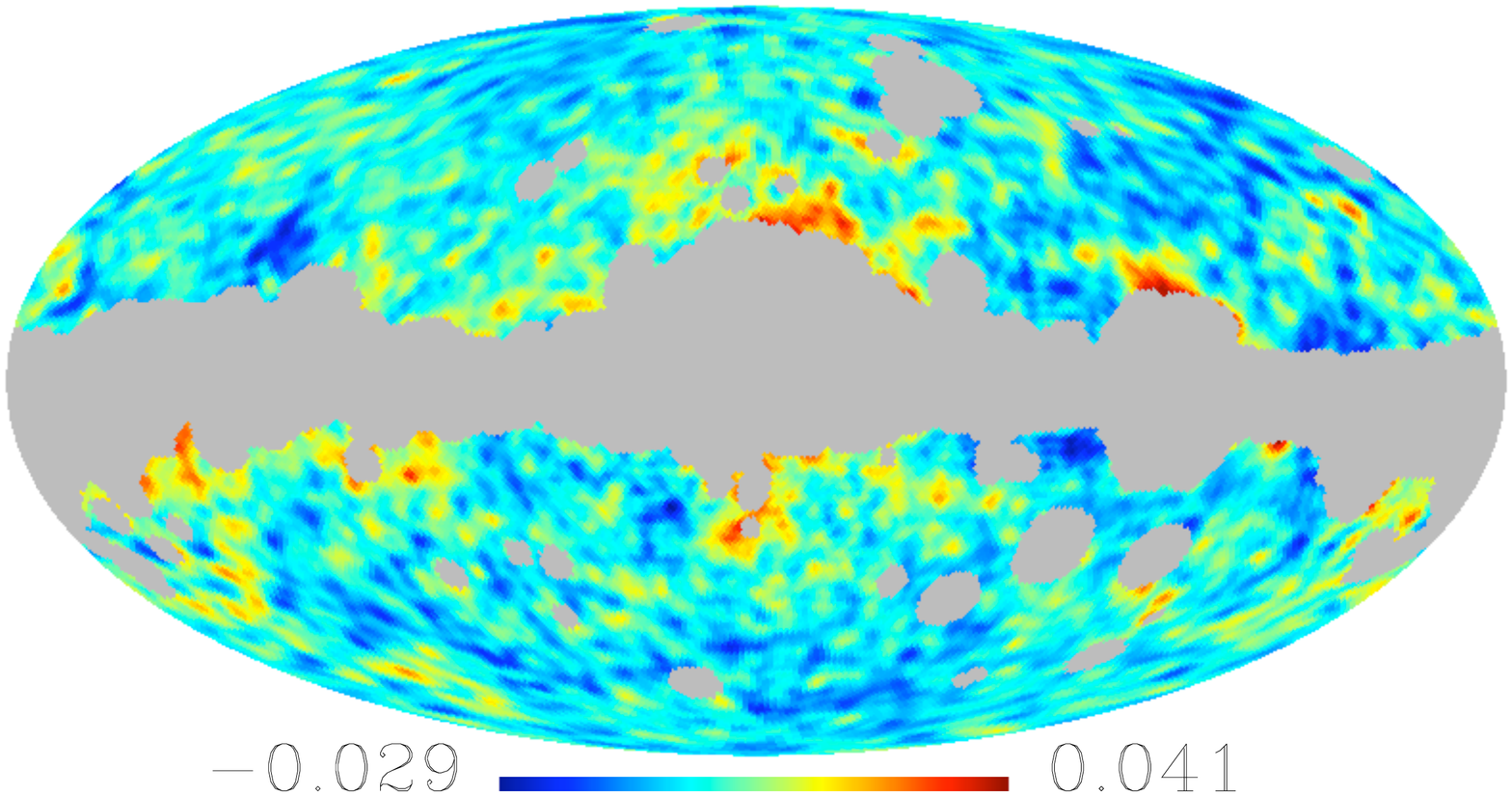}
\caption{Residual emission, in units of mK antenna temperature, in the K (top), Ka (middle) and Q (bottom) frequency bands, when using the Gibbs (left) and ILC (right) CMB estimators.}
\label{fig:diff_all_3temp}
\end{center}
\end{figure}
In fig.~\ref{fig:diff_all_3temp} we see the result of this procedure, which shows the residual emission in the K (top), Ka (middle) and Q (bottom) WMAP bands when using the Gibbs (left) and ILC (right) CMB estimators. (We do not display residual maps for the V and W bands since the large experimental errors $\sigma_{i,j}$ heavily suppress the variance at these frequencies.) The overall $\chi^2_{\rm red.}$ of the fit is equal to 7.85 (11.05) for the Gibbs (ILC) analysis. 
To further gauge the significance of the residual emission we calculate the following parameter
\be
\zeta_{i}^2=\frac{1}{N_p}\sum\limits_{j=1}^{N_p}\zeta_{i,j}^2=\frac{1}{N_p}\sum\limits_{j=1}^{N_p}\frac{\left(x_{i,j}-\langle x\rangle_{i,j} \right)^2}{\sigma_{i,j}^2},
\label{eq:zeta_i:ch3}
\ee
where $x_{i,j}$ and $\sigma_{i,j}$ are the respective values of the $j$th pixel in the $i$th frequency band associated with the residual map and CMB $1\sigma$ error map associated with that band. $\langle x \rangle_{i,j}=0$ is the expectation value of $x_{i,j}$, since we expect, as a first approximation, that we will obtain a perfect fit to the data. Therefore, looking at the form of (\ref{eq:zeta_i:ch3}) we can say that regions of statistically significant residual signal possess values $\zeta_{i,j}^2\gtrsim1$.
\begin{figure}[h]
\vspace{0.5cm}
\begin{center}
\includegraphics[scale=0.17, bb= 100 200 510 700, angle=90]{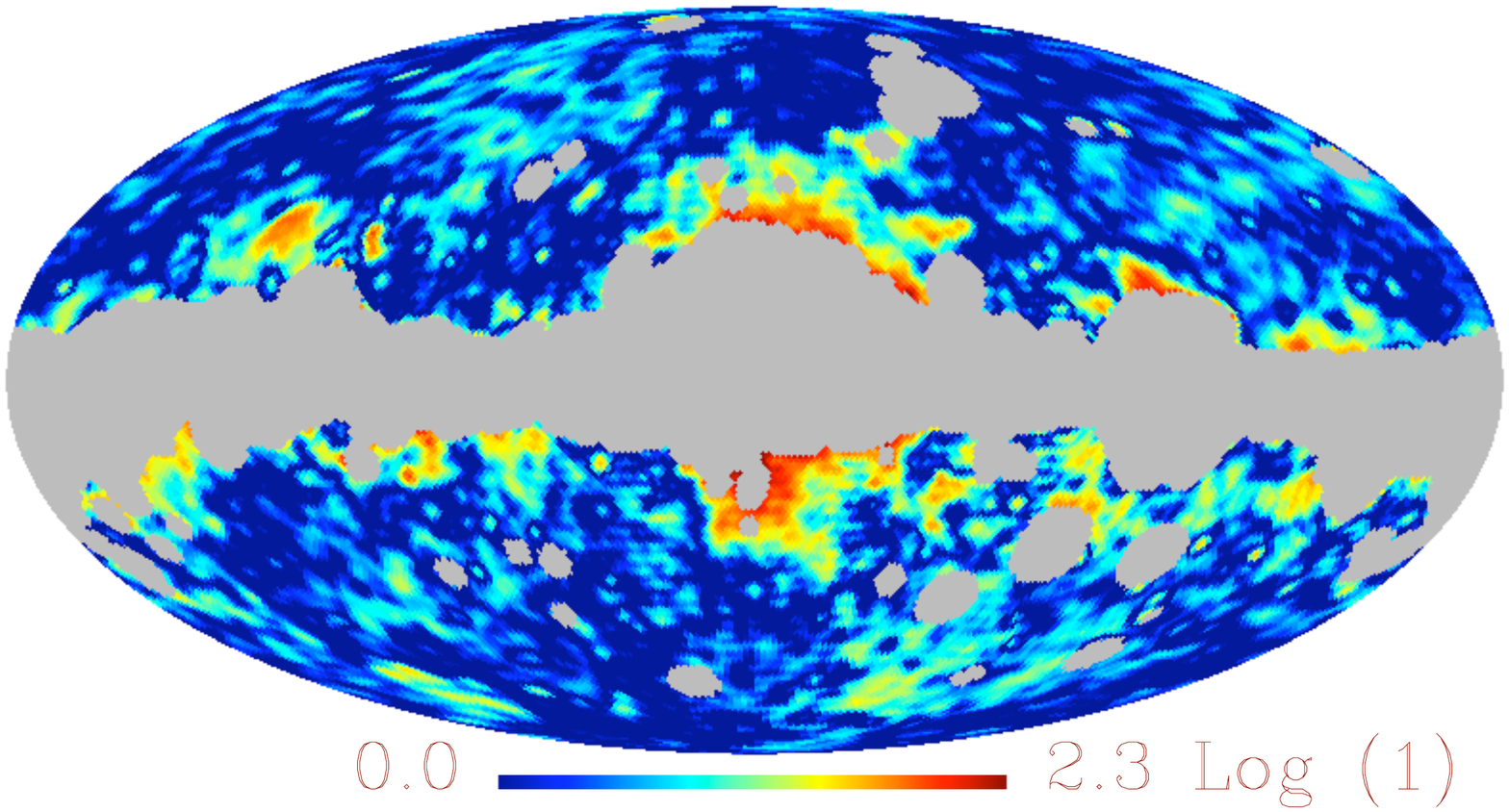}
\hspace{0.1cm}
\includegraphics[scale=0.17, bb= 100 150 510 900, angle=90]{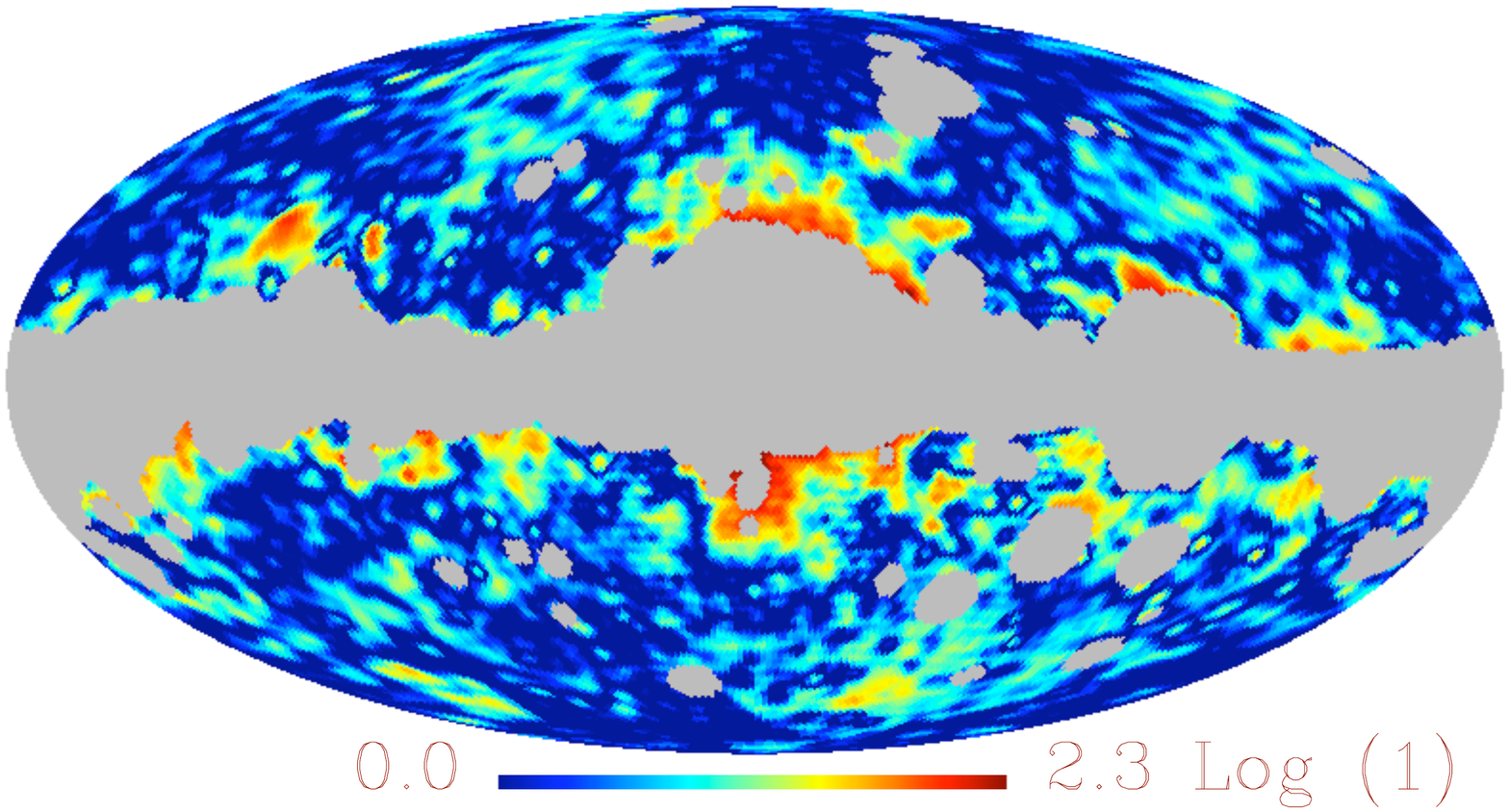}
\includegraphics[scale=0.17, bb= 100 200 510 700, angle=90]{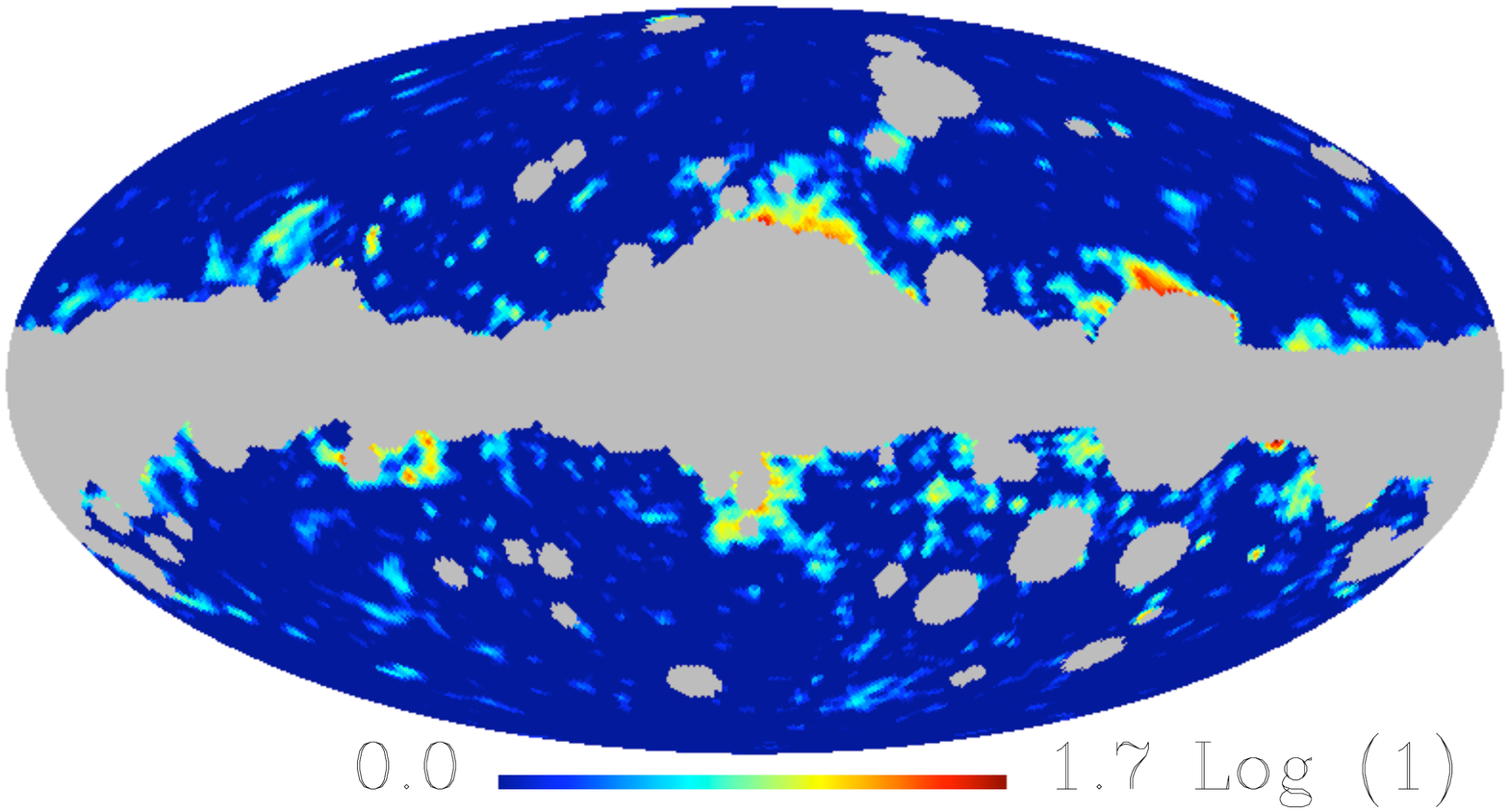}
\hspace{0.1cm}
\includegraphics[scale=0.17, bb= 100 150 510 900, angle=90]{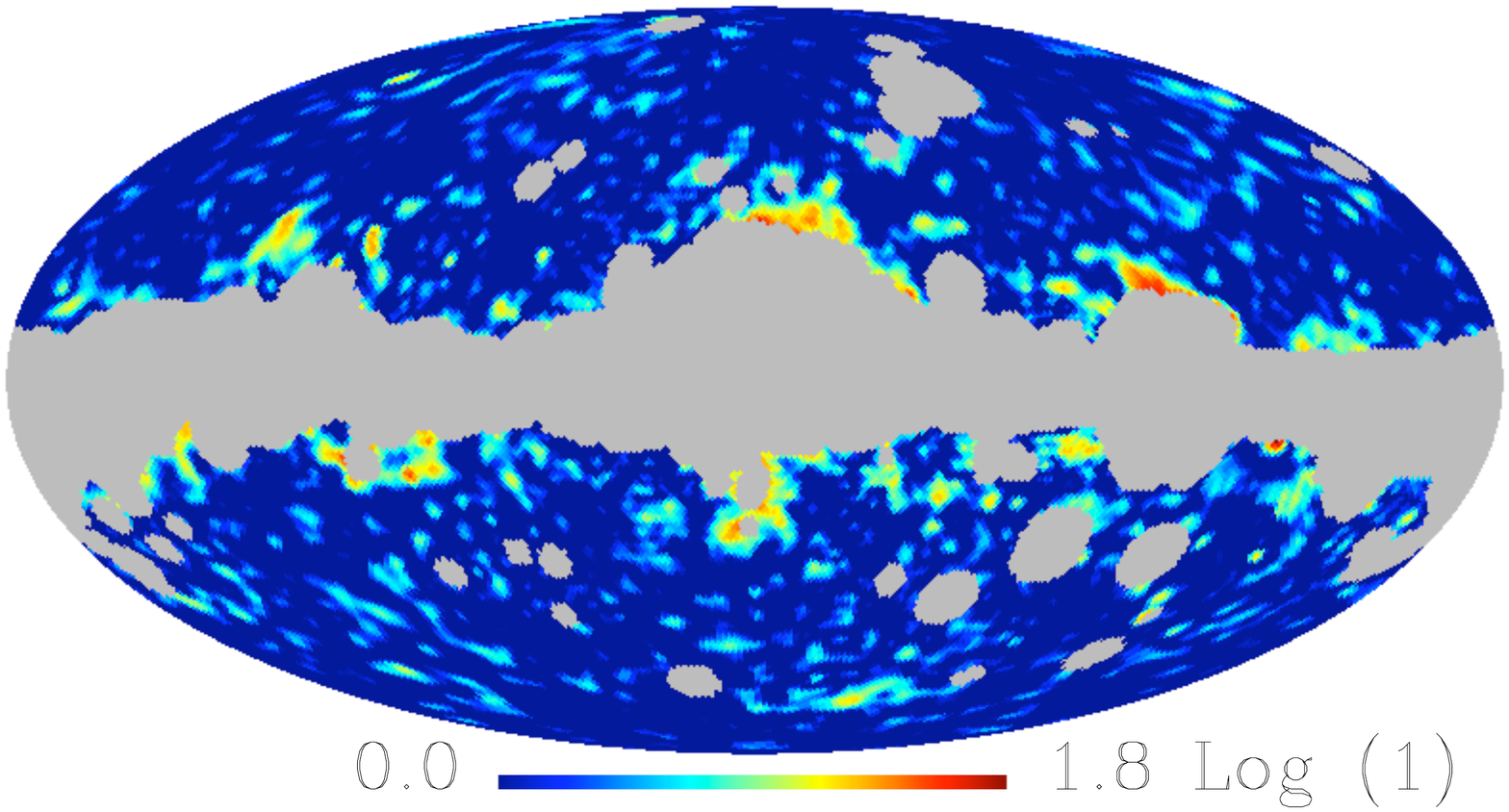}
\includegraphics[scale=0.17, bb= 100 200 510 700, angle=90]{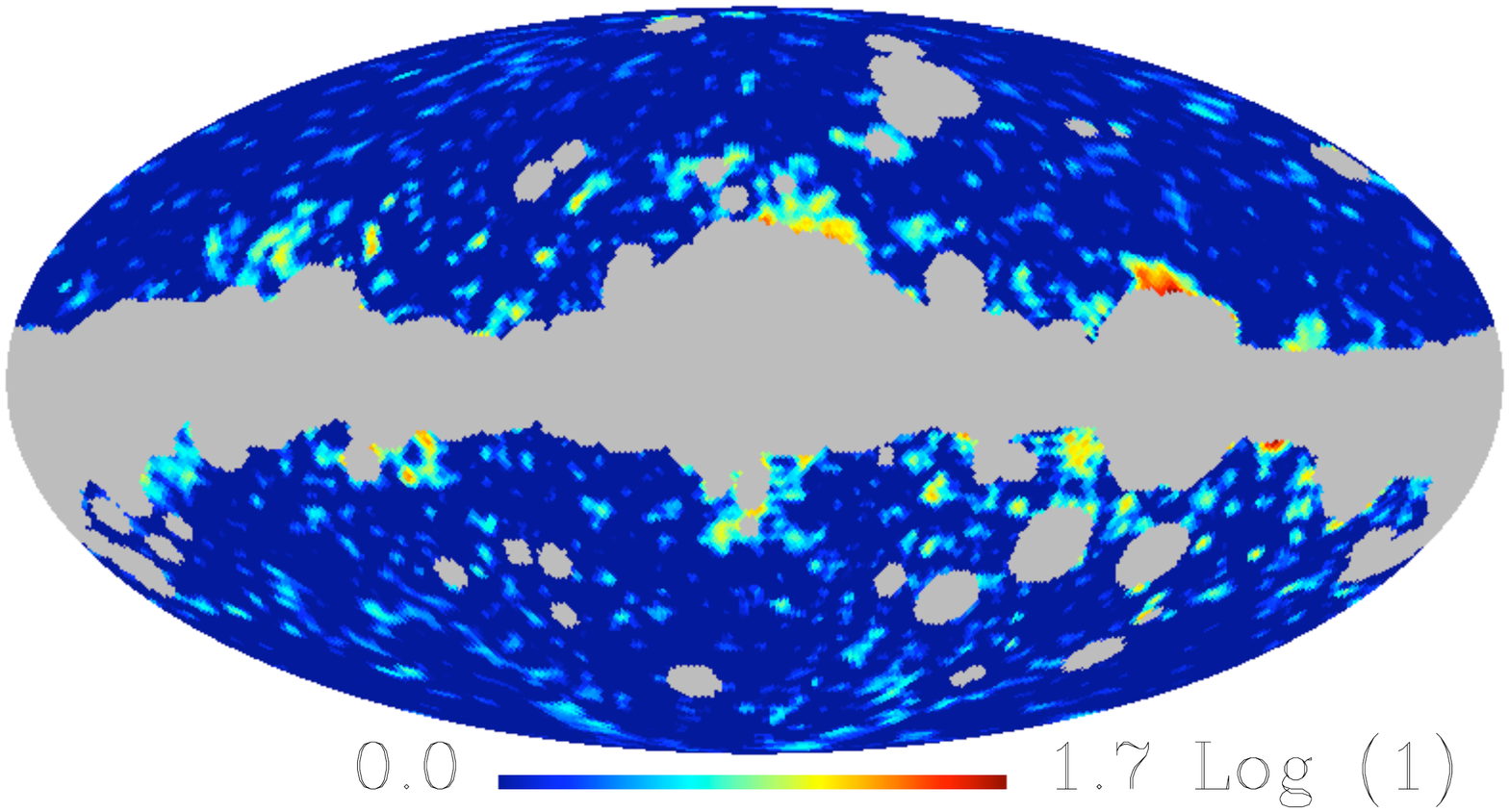}
\hspace{0.1cm}
\includegraphics[scale=0.17, bb= 100 150 510 900, angle=90]{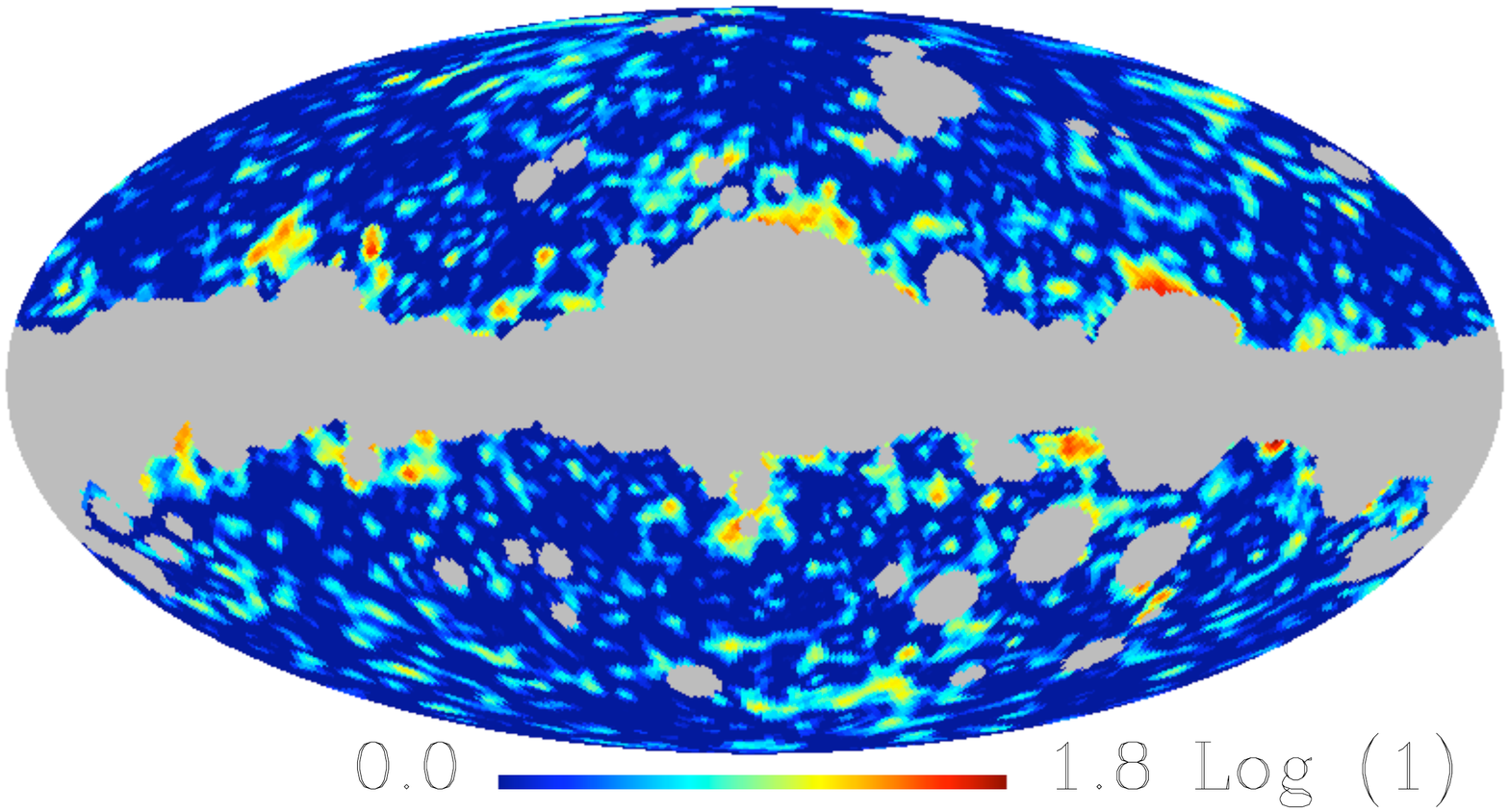}
\caption{Maps of the parameter $\zeta_{i,j}^2$ in the K (top), Ka (middle) and Q (bottom) frequency bands, when using the Gibbs (left) and ILC (right) CMB estimators, where we truncate the maps using a minimum of $\zeta_{i,j}^2=1$.}
\label{fig:zeta_all_3temp}
\end{center}
\end{figure}

In fig.~\ref{fig:zeta_all_3temp} we display maps of $\zeta_{i,j}^2$ in the K (top), Ka (middle) and Q (bottom) bands, for both Gibbs (left) and ILC (right) analyses, truncating the maps with a minimum of $\zeta_{i,j}^2$=1, i.e. only displaying regions of statistically significant residual emission. The values of $\zeta^2_{i}$ corresponding to the residual maps in each of the five frequency bands are:  5.54 (6.59), 0.88 (1.45), 1.08 (2.12), 0.089 (0.174) and 0.028 (0.271) for the K, Ka, Q, V and W bands respectively, for the Gibbs (ILC) analysis. The $1\sigma$ errors associated with these values are all of order $10^{-5}$. Hence it is clear that a statistically significant (i.e. $\zeta_{i}^2\gtrsim1$) residual emission remains, at least for the K band, which appears to be concentrated mostly around the GC. In fact, the respective values of $\zeta_{i}^2$ calculated solely for the sky region within 50$^{\circ}$ of the GC increase to 14.69 (16.59), 1.65 (2.42), 1.60 (2.84), 0.11 (0.20) and 0.027 (0.30). The $1\sigma$ errors associated with these values are all of the order $10^{-2}$ or less. The errors associated with the results presented throughout this paper are $1\sigma$ errors (unless stated otherwise). They were calculated using a random sample of 1000 realizations of the CMB data maps, constructed using the average CMB-removed WMAP data and the 1$\sigma$ error maps $\mathbf{\sigma}$.

Using the elements of the solution vector ${\bf a}$ we can calculate the spectral dependence of the various foreground components. 
\begin{figure}[h]
\begin{center}
\vspace{1.0cm}
\includegraphics[width=90mm, keepaspectratio, clip]{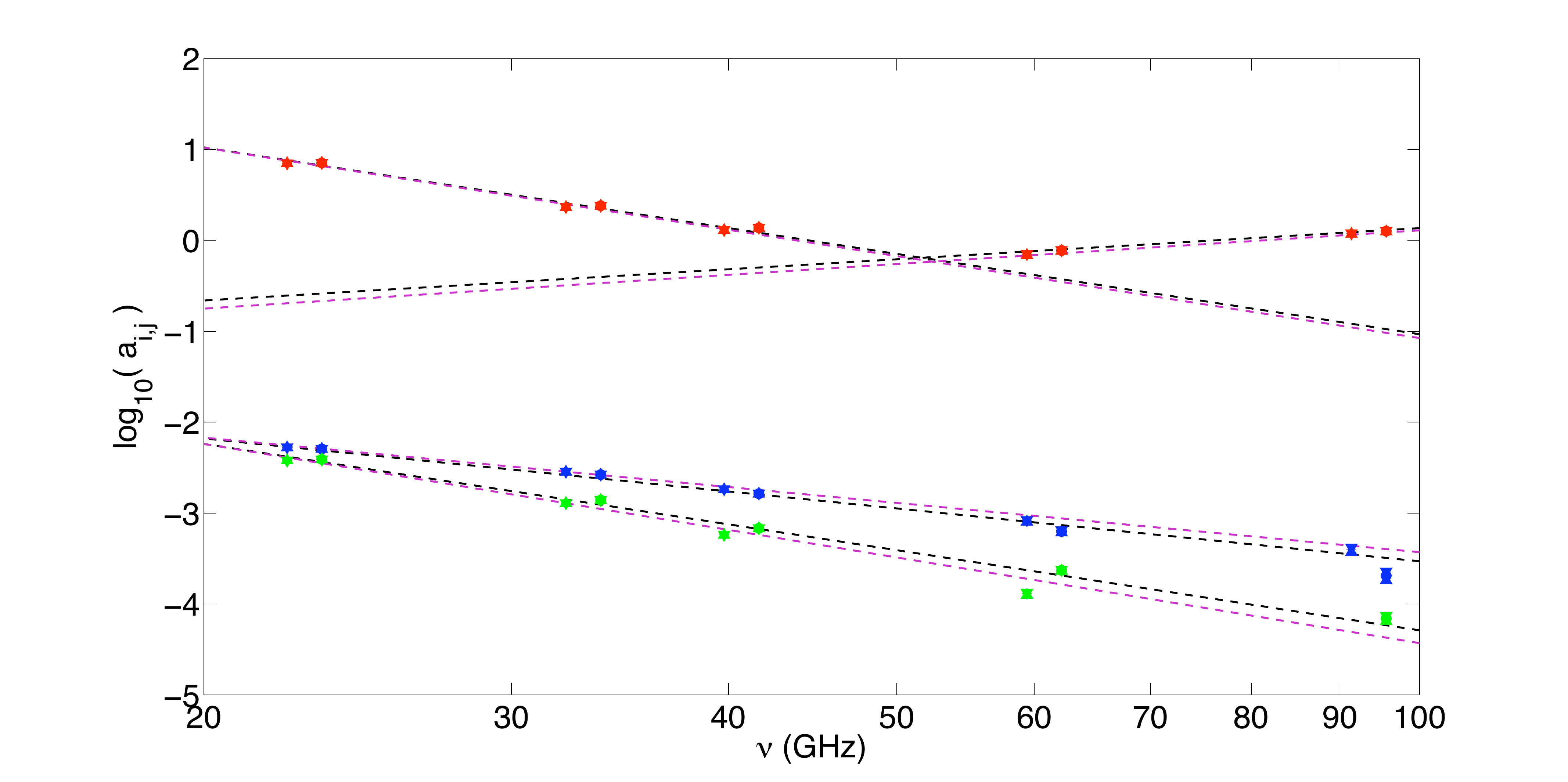}
\caption{Spectral dependences of best fitted free-free (blue markers), dust (red markers) and synchrotron emission (green markers) foreground components resulting from our fit when using the Gibbs (circles) and ILC (crosses) CMB estimators. Also plotted are several lines of best fit associated with the best fit spectral indices for the data points (dashed black (Gibbs) and dashed magenta (ILC) ) associated with free-free and synchrotron components, and also for dust emission in the K-Ka and V-W dust frequency bands. For clarity, we have slightly displaced the frequency of the two sets of data markers associated the Gibbs and ILC CMB estimators by $\Delta{\rm log}_{10}(\nu /{\rm GHz})$=+0.5 and $-0.5$ respectively.}
\label{fig:spec3temp:ch3}
\end{center}
\end{figure}
In fig.~\ref{fig:spec3temp:ch3} we plot the spectral dependences of the free-free (blue markers), dust (red markers) and synchrotron emission (green markers) resulting from our fit using the Gibbs (circles) and ILC (crosses) CMB estimators. We obtain best fit values of spectral index $\alpha$ equal to -1.93 (-1.80) and -2.93 (-3.13) for the free-free and synchrotron emission components respectively, when using the Gibbs (ILC) estimators, where we assume that the foreground intensity $I_{\nu}$ scales with frequency as $I_{\nu}\propto\nu^{\alpha+2}$. These results are consistent with those determined in previous studies discussed in \S\,\ref{subsec:fftemp:ch3} and \S\,\ref{subsec:synctemp:ch3}. To further illustrate this we have plotted lines of best fit with these gradients (dashed black (Gibbs) and dashed magenta (ILC) ). For clarity, we have slightly displaced the frequency of the two sets of data markers associated the Gibbs and ILC CMB estimators by $\Delta{\rm log}_{10}(\nu /{\rm GHz})$=+0.5 and $-0.5$ respectively.

Unlike free-free and synchrotron emission, the emission from dust does not have a constant spectral index, but varies from -2.93 (-2.99) between the K and Ka bands to +1.14 (+1.23) between the V and W bands when using the Gibbs (ILC) analysis, with a minimum intensity occurring around 50\,GHz. Again we illustrate this by plotting lines of best-fit with these gradients. Such results are consistent with the notion that there are two components to dust emission, dominating at the opposite extremities of the WMAP frequency range. It has been suggested that emission from both spinning dust, dominating in the K, Ka and Q bands with a large negative index of approximately -2.85, and thermal dust, dominating in the W band with a positive spectral index of approximately +1.7, may be consistent with the observed spectral behaviour of the dust emission (see e.g.~\cite{ Davies2006:ch3, Finkbeiner2007b:ch3}), however we do not to pursue the investigation of these topics further.

We now briefly consider the correlation between the residual haze and the various foreground components. We define the elements of correlation matrix, $\phi_{{\bf X},{\bf Y}}$, involving foreground template ${\bf X}$ and residual or foreground template ${\bf Y}$ (with elements $x_i$ and $y_i$ respectively) as
\be
\phi_{{\bf X},{\bf Y}}=\frac{{\langle\left[x_i-\langle x \rangle\right]\left[y_i-\langle y \rangle\right]\rangle}}
{\sqrt{ \langle\left[y_i-\langle y \rangle\right]^2\rangle\langle\left[y_i-\langle y \rangle\right]^2\rangle  }}.
\ee
In table~\ref{tbl:corrmat:ch3} we display the correlation matrix involving each of the three foreground templates and the K-band residual map when using the Gibbs CMB estimator (the correlation matrix when using the ILC estimator is extremely similar).  One can clearly observe that the haze emission in the K-band is significantly more correlated with synchrotron emission (0.48) than either the dust (0.19) or free-free (0.11) emission. We illustrate this correlation in the following section where we demonstrate how we can substantially improve the quality of the fit to the data by allowing the spectral dependence of the soft synchrotron emission for regions close to the GC to vary independently from that associated with regions far away from the GC.
\begin{table}
\begin{center}
\begin{tabular}{|c|c|c|c|c|}\hline
&~Free-free~&~~~Dust~~~&~Synchrotron~&~~~Haze~~~\\
\hline
~~Free-free~~&1.00&0.48&0.08&0.11\\
\hline
~~~~Dust~~~~~& &1.00&0.44&0.19\\
\hline
~Synchrotron~& & &1.00&0.48\\
\hline
Haze& & & &1.00\\
\hline
\end{tabular}
\caption{Correlation matrix involving the free-free, dust and synchrotron foreground templates together with the map of residual emission in the K-band following the 3-template fit.}
\label{tbl:corrmat:ch3}
\end{center}
\end{table}

\section{Fit using multiple synchrotron templates}

In this section we investigate whether the significant correlation between the residual haze and the soft synchrotron galactic foreground emission demonstrated in section \S\,\ref{sec:3temp} can be explained by a subtle {\it spatial} variation in the spectral dependence of soft synchrotron galactic foreground emission.  Here, we adopt a simple, yet illustrative model to allow for such spatial variation by splitting the synchrotron template into two regions (one central and one peripheral to the GC), and allow the two to have different spectral indices. We propose, for simplicity, that the boundary separating the two regions should be symmetrical around the GC, corresponding to all lines of sight with inclination $\theta$ relative to the GC.
\begin{figure}[h]
\begin{center}
\includegraphics[scale=0.25, bb= 100 100 510 720, angle=90]{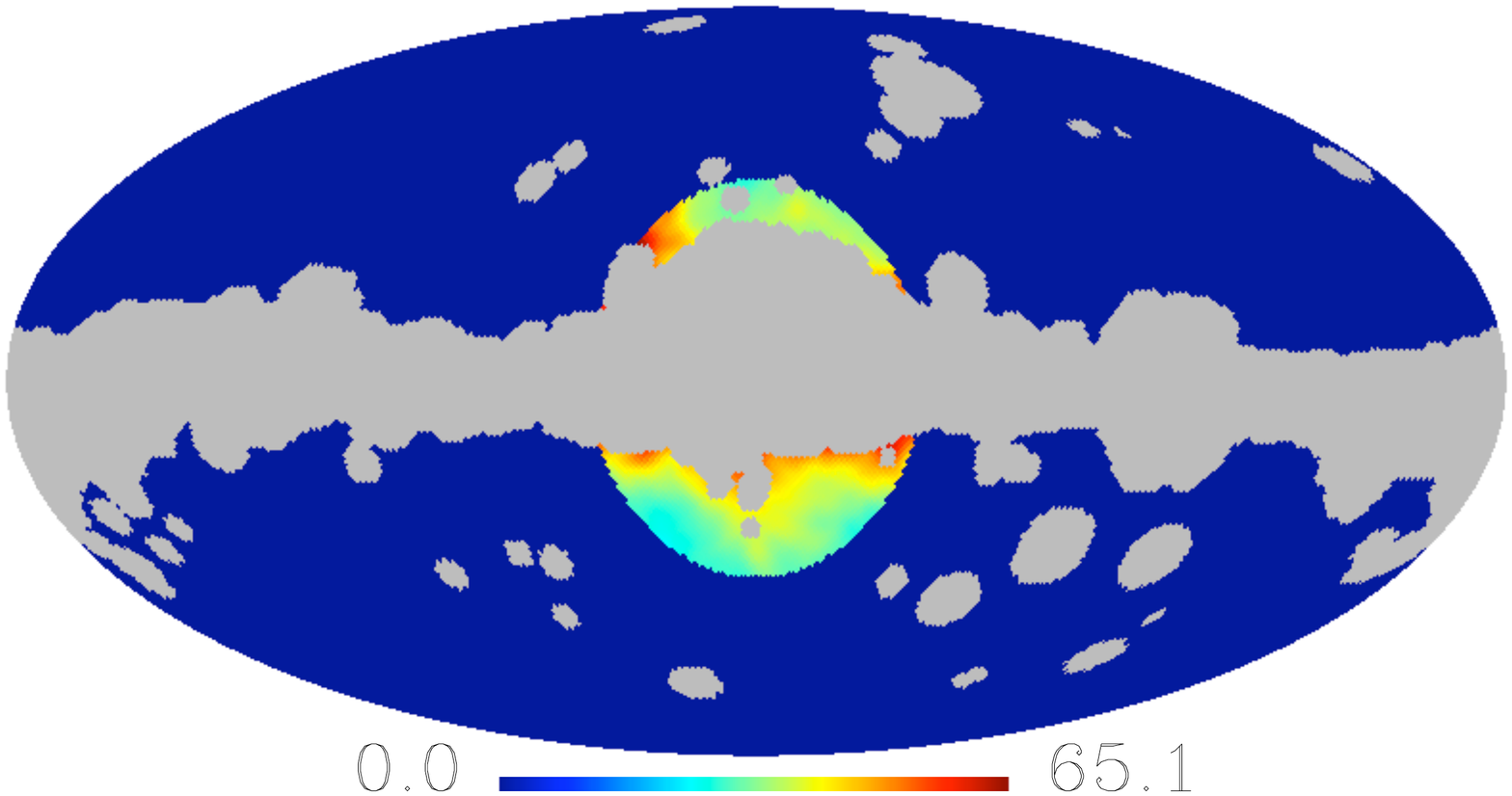}
\hspace{1.1cm}
\includegraphics[scale=0.25, bb= 100 100 510 720, angle=90]{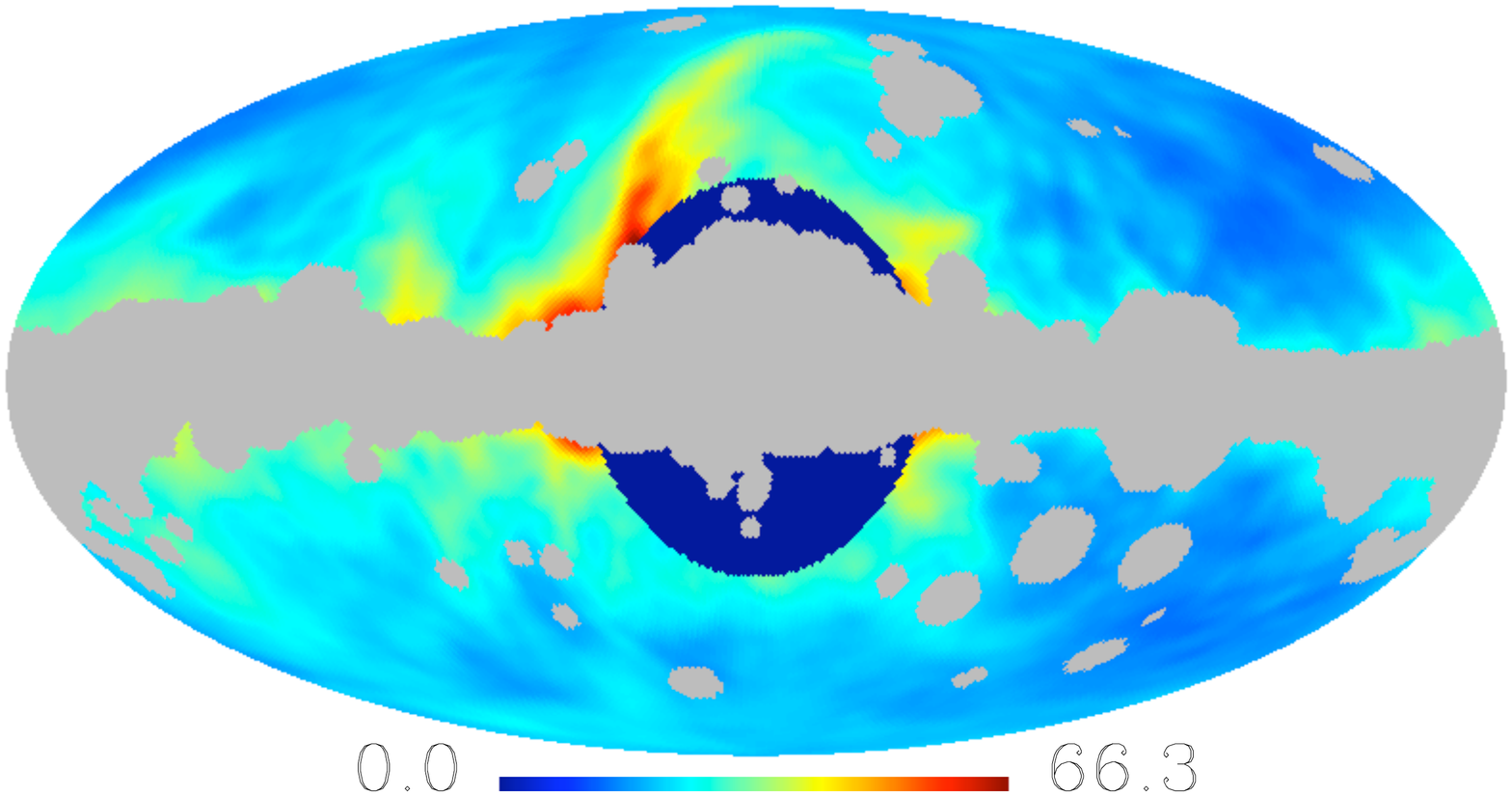}
\caption{Central (top) and peripheral (bottom) templates of soft synchrotron galactic foreground emission for $\theta=50^{\circ}$, presented in units $[ T(K) ]$, where $T$ is antenna temperature.}
\label{fig:sync_in_out:ch3}
\end{center}
\end{figure}

In fig.~\ref{fig:sync_in_out:ch3} we illustrate an example case where the Haslam {\it et~al.} synchrotron template is segregated into central and peripheral regions according to the prescription described above when using $\theta=50^{\circ}$. 
We can clearly see that if the majority of the residual emission is localised within the central region and is dominated by synchrotron radiation from an independent, possibly exotic, source (e.g. such as dark matter), then we expect that the spectral indices describing the synchrotron emission in the central and peripheral regions to be significantly different.

We formulate our conclusions based upon results from a four-template multi-linear fit to the CMB-subtracted WMAP data, involving our free-free, dust and two new synchrotron templates, defined by an angle $\theta$ which minimises the value of $\chi^2_{\rm red.}(\theta)$ associated with the fit.
\begin{figure}
\begin{center}
\includegraphics[width=90mm, keepaspectratio, clip]{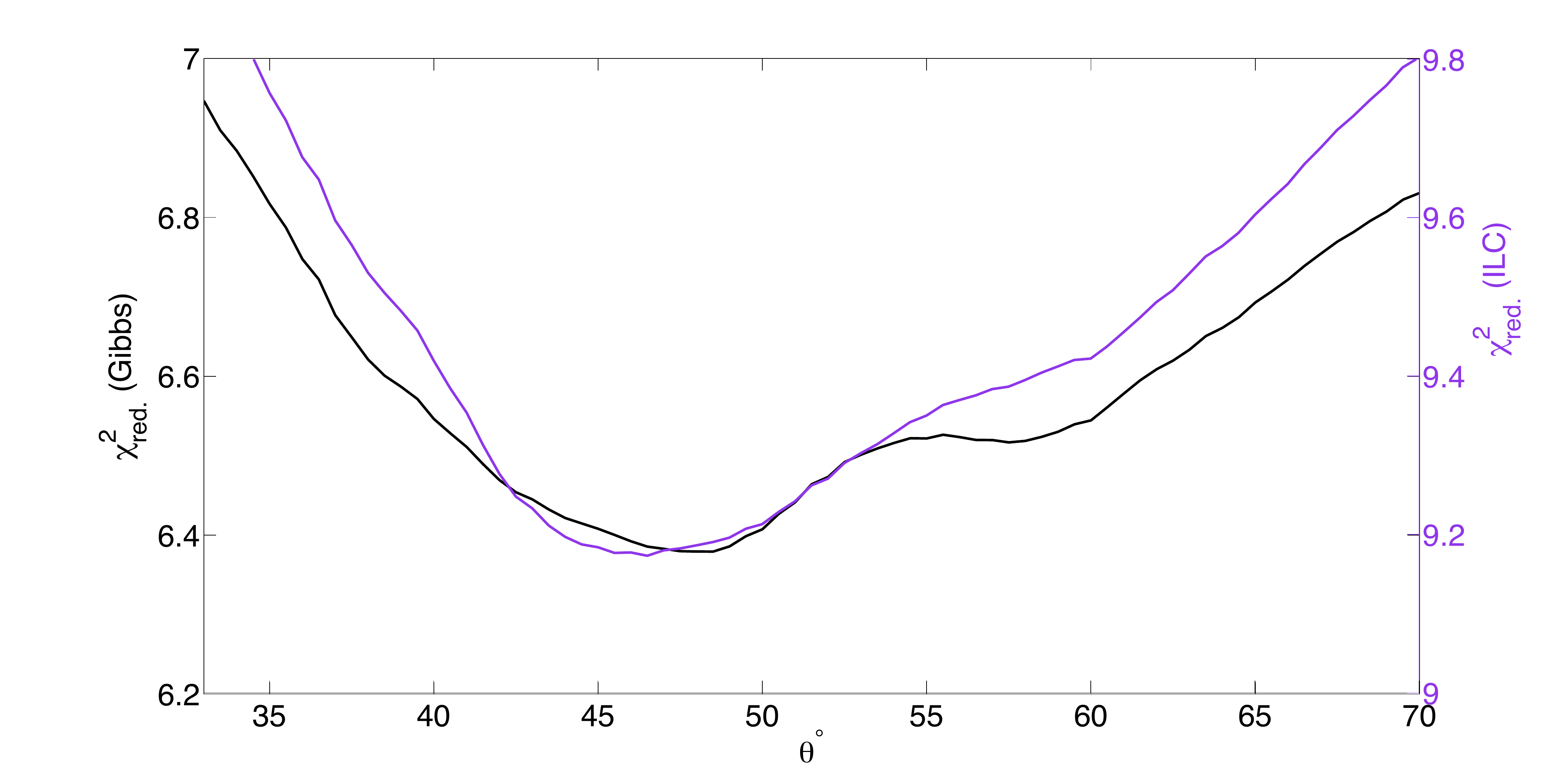}
\caption{$\chi^2_{\rm red.}$ verses $\theta$ for 4-template multi-linear fits to the CMB-subtracted WMAP data, involving our free-free, dust, central and peripheral synchrotron templates when using the Gibbs (black) and ILC (magenta) CMB estimators.}
\label{fig:chisqvstheta:ch3}
\end{center}
\end{figure}
\begin{figure}[h]
\begin{center}
\vspace{1cm}
\includegraphics[scale=0.17, bb= 100 200 510 700, angle=90]{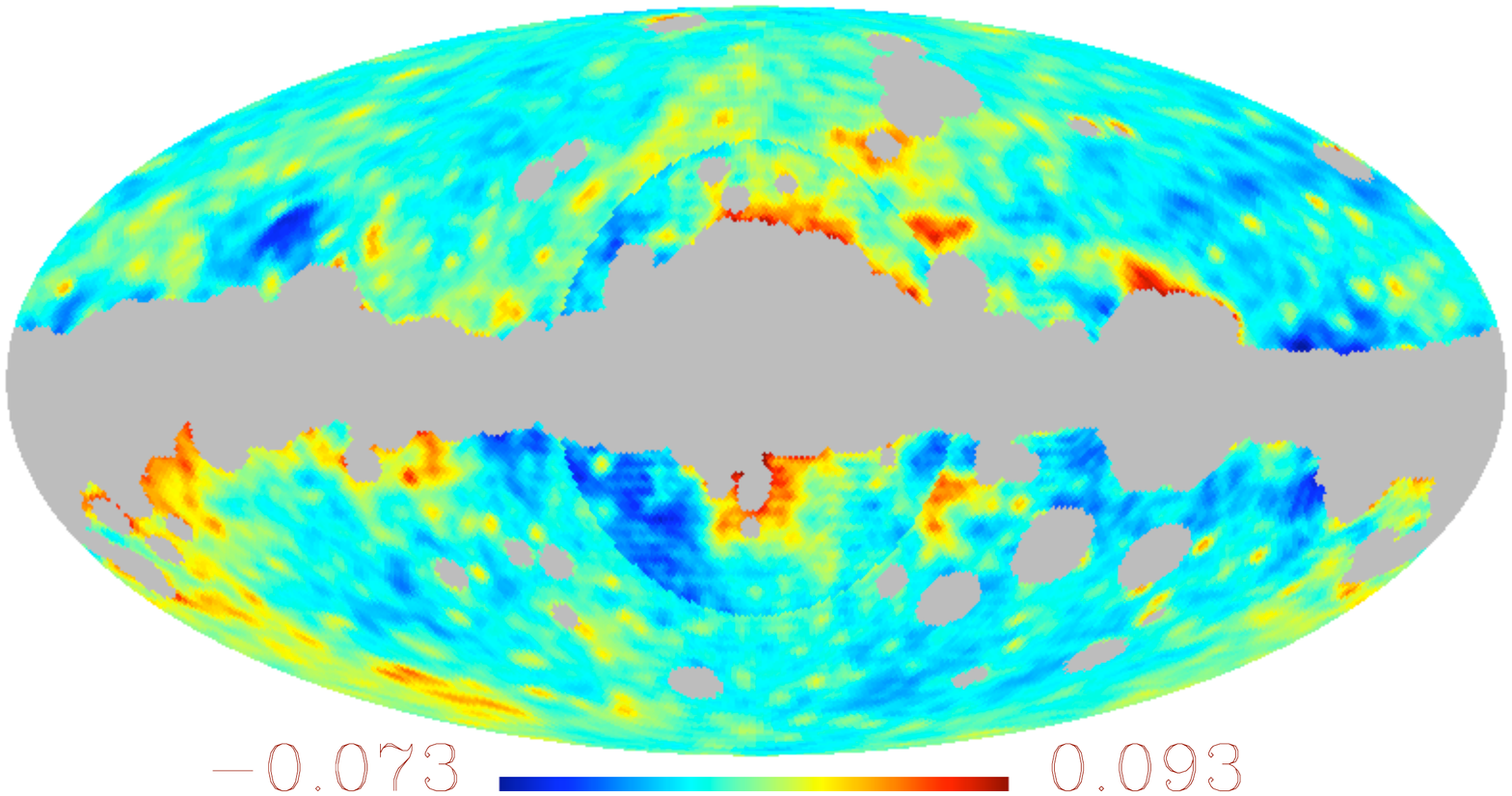}
\hspace{1.1cm}
\includegraphics[scale=0.17, bb= 100 200 510 720, angle=90]{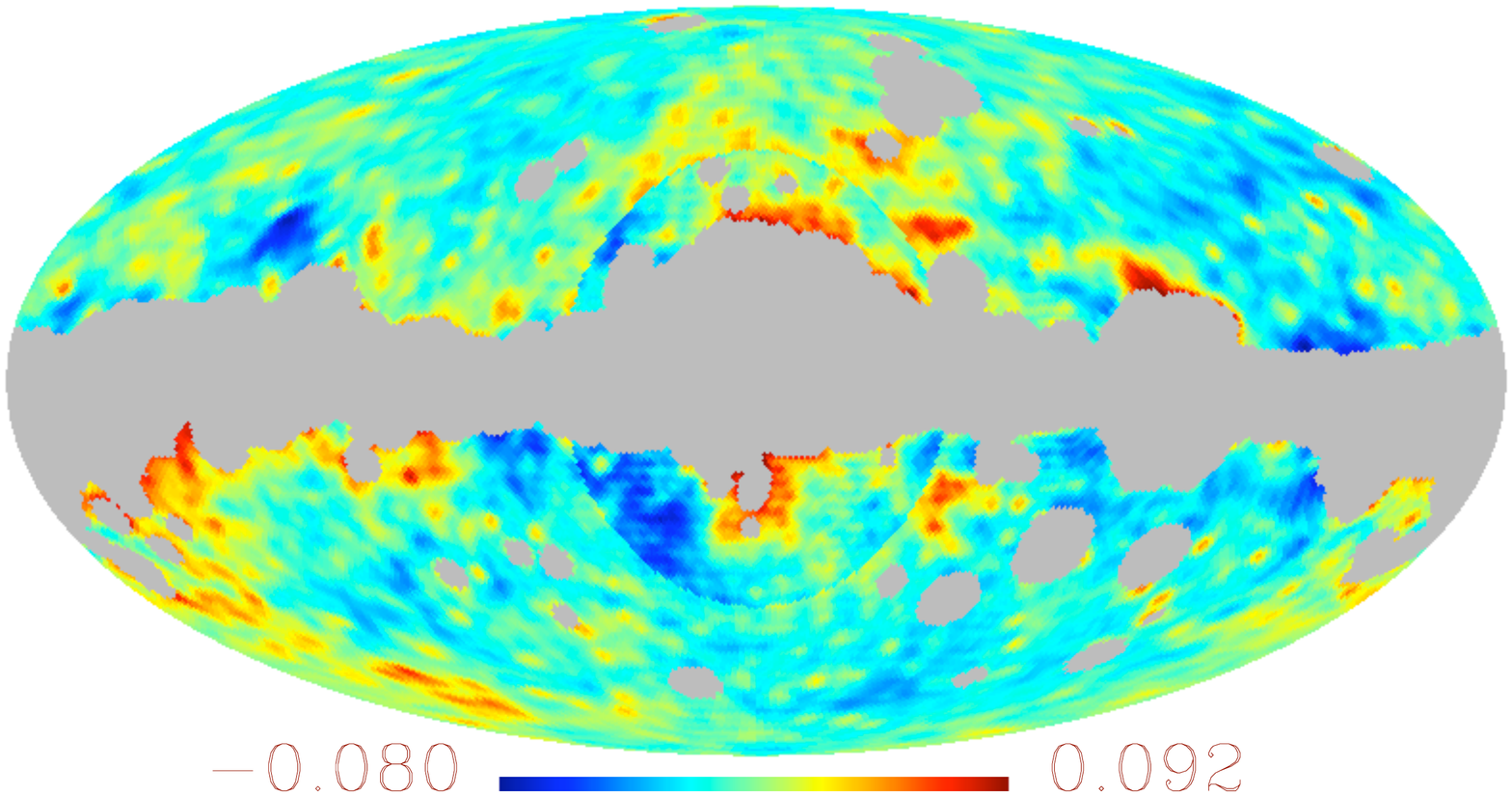}
\includegraphics[scale=0.17, bb= 100 200 510 700, angle=90]{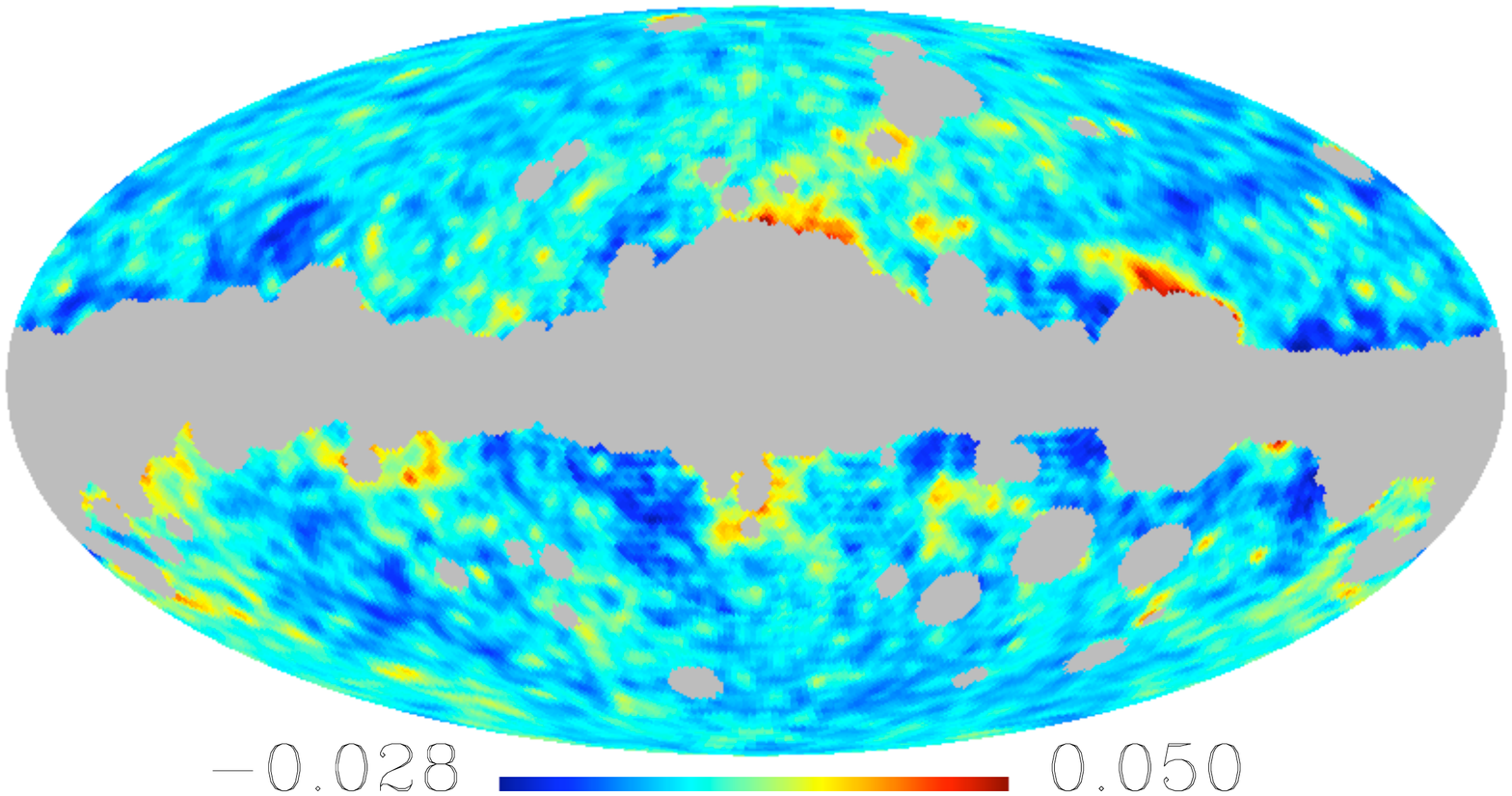}
\hspace{1.1cm}
\includegraphics[scale=0.17, bb= 100 200 510 720, angle=90]{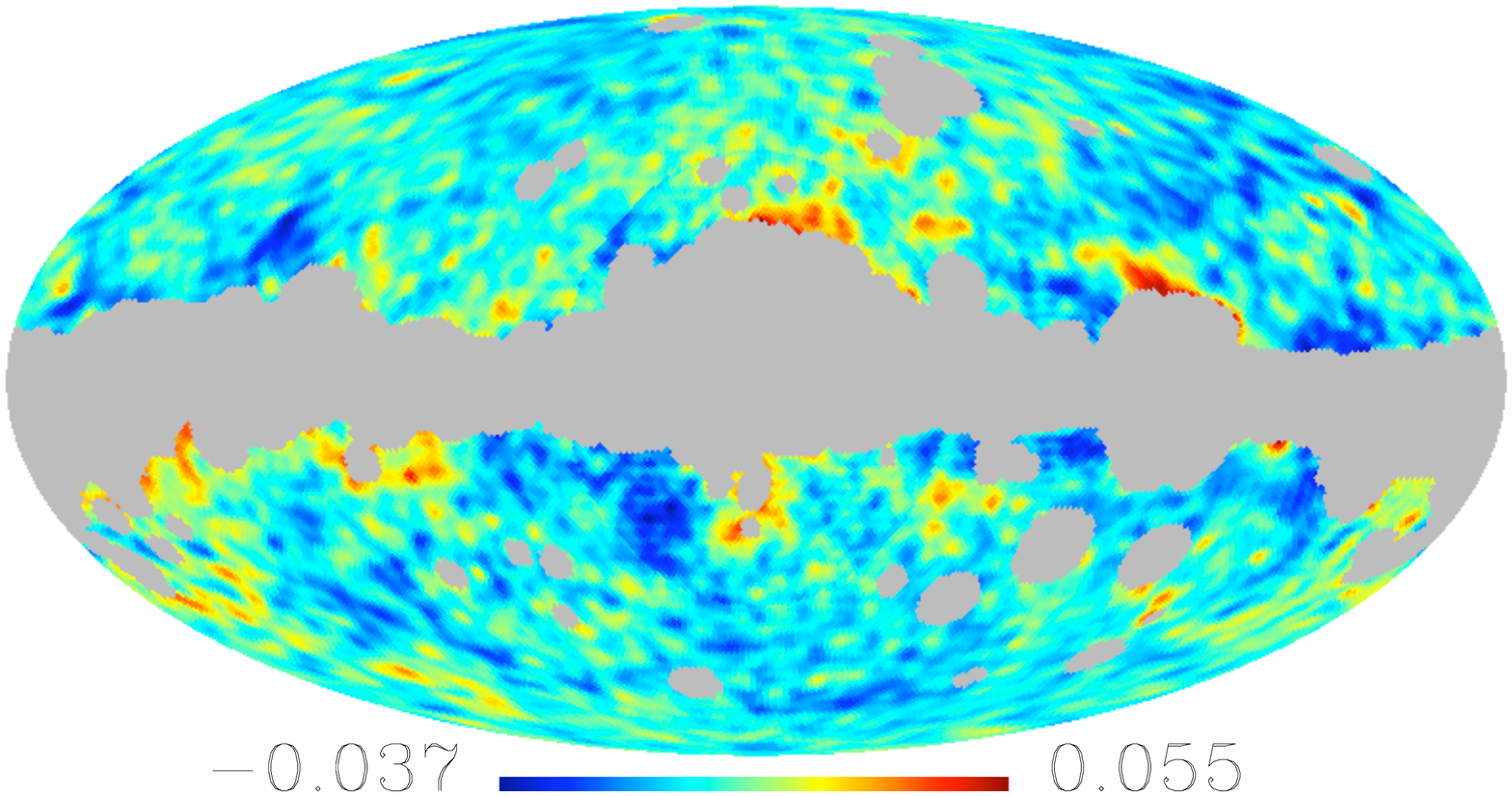}
\includegraphics[scale=0.17, bb= 100 200 510 700, angle=90]{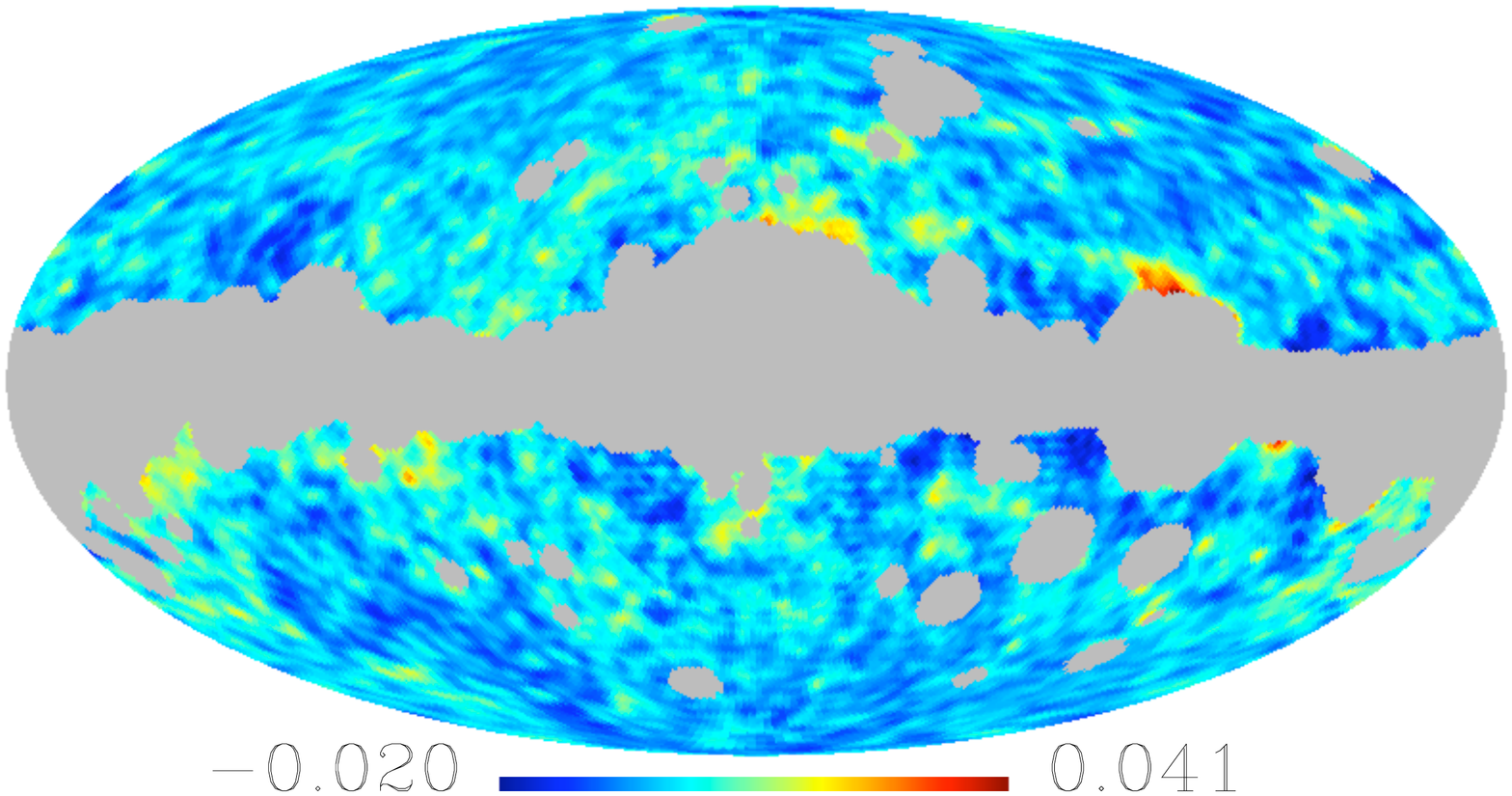}
\hspace{1.1cm}
\includegraphics[scale=0.17, bb= 100 200 510 720, angle=90]{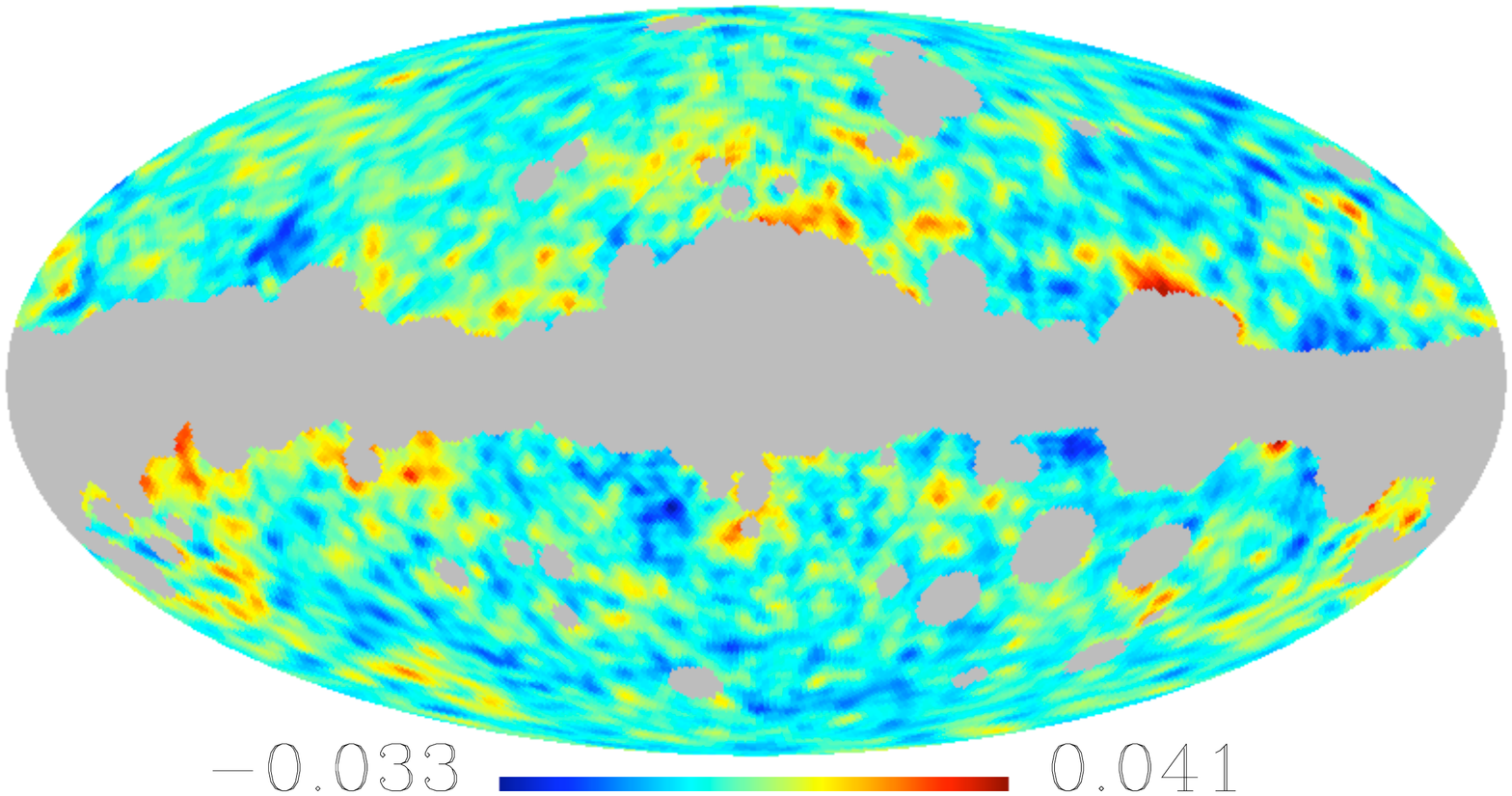}
\includegraphics[scale=0.17, bb= 100 200 510 700, angle=90]{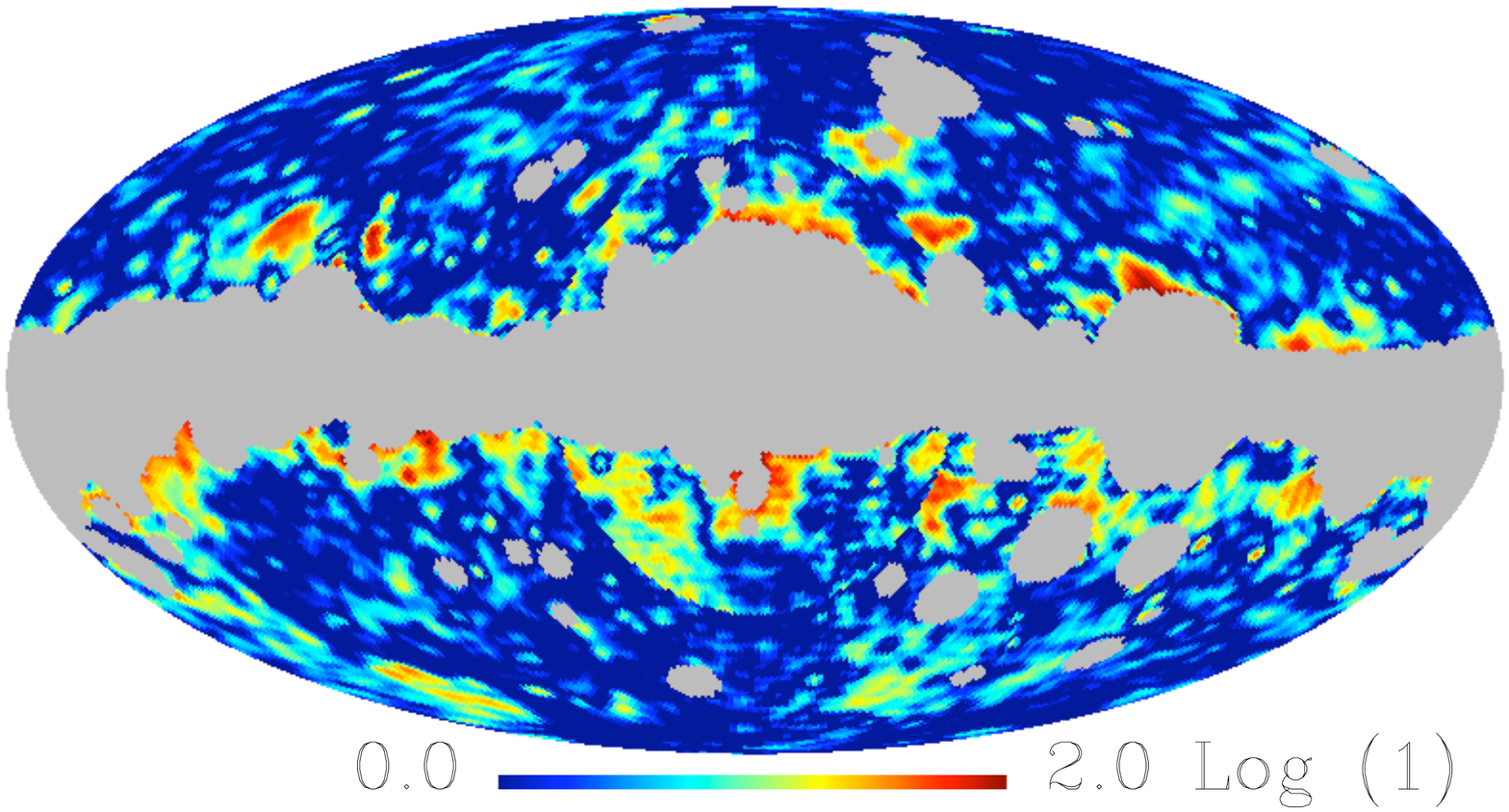}
\hspace{1.1cm}
\includegraphics[scale=0.17, bb= 100 200 510 720, angle=90]{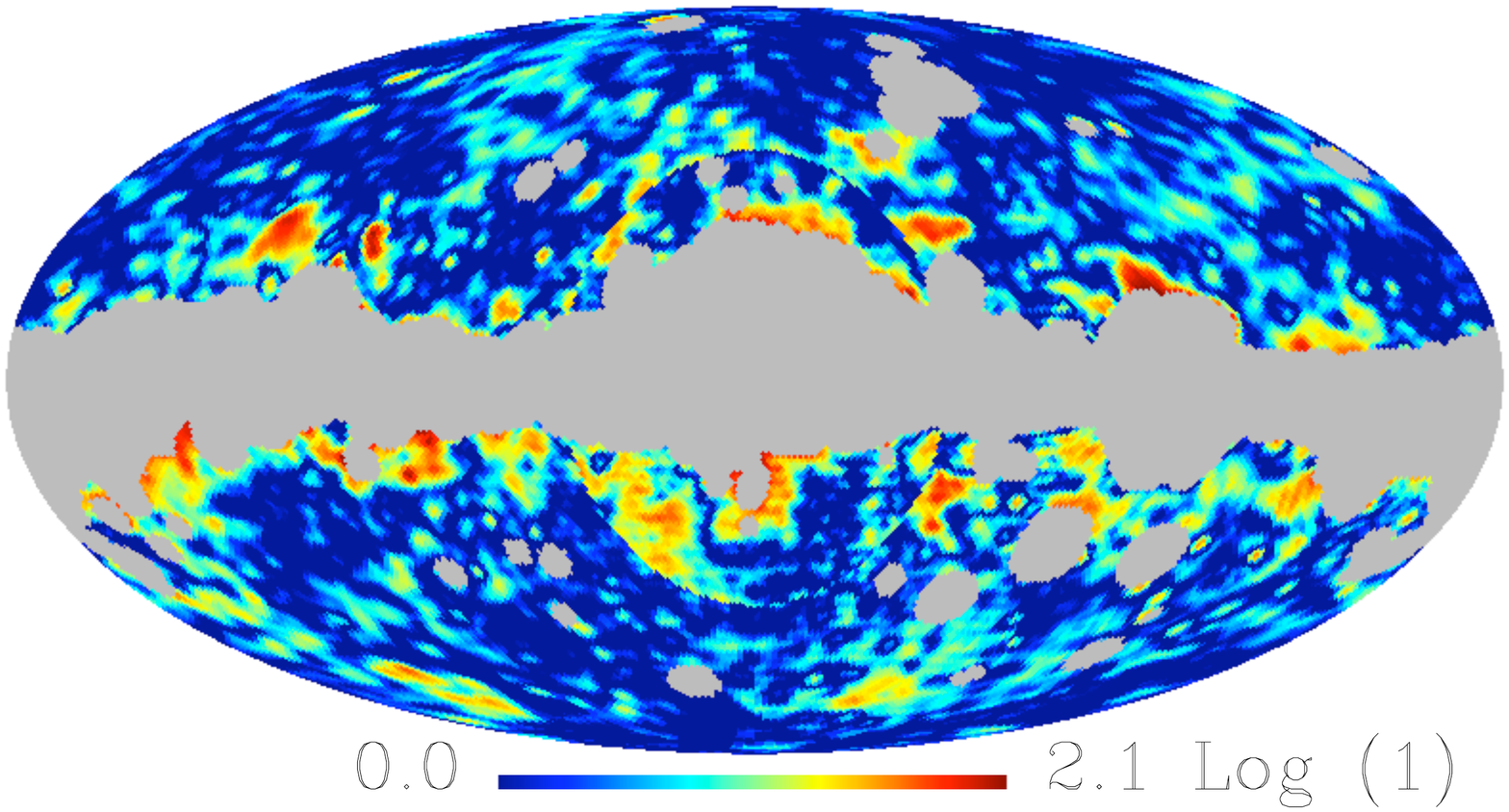}
\includegraphics[scale=0.17, bb= 100 200 510 700, angle=90]{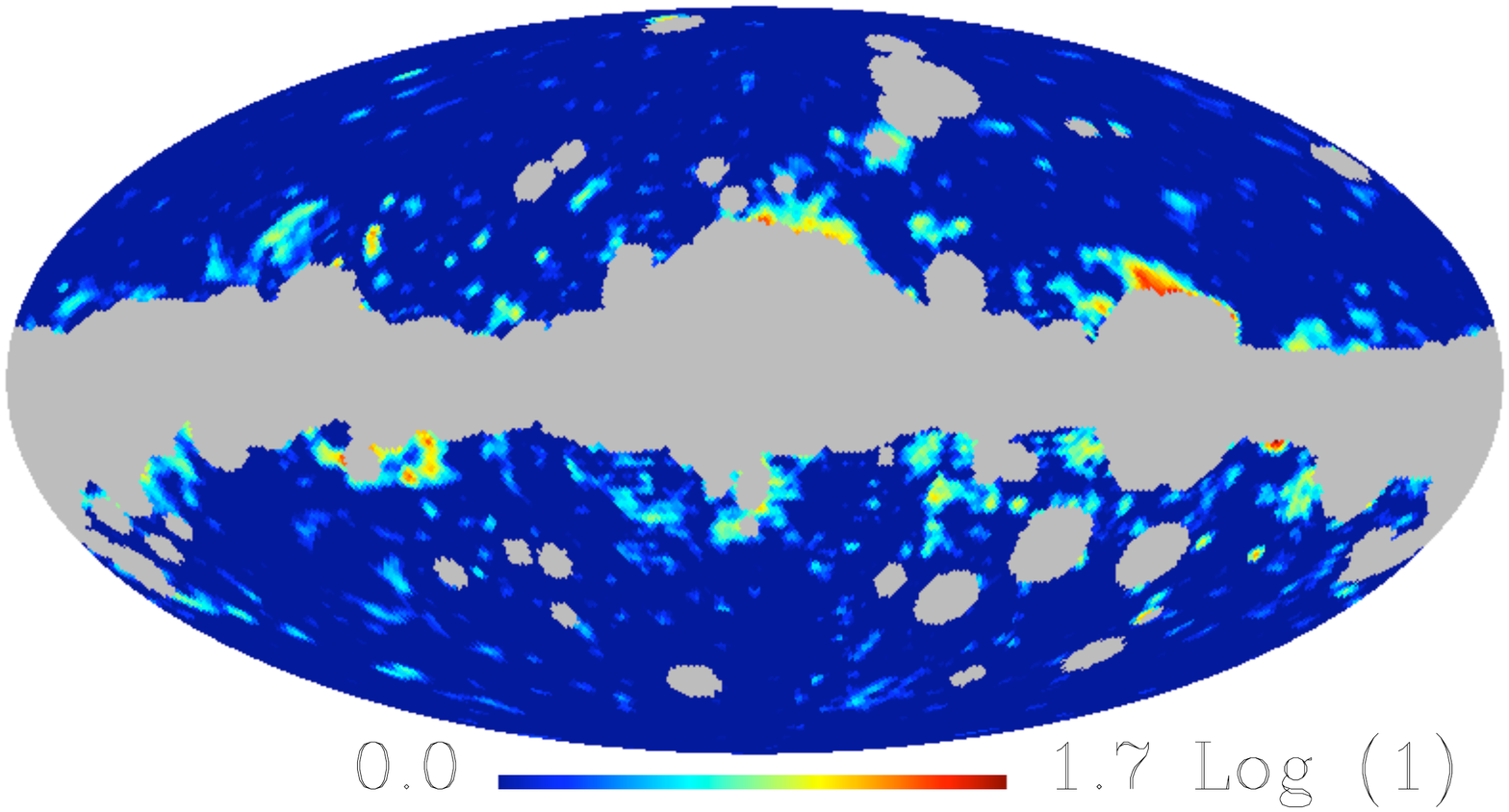}
\hspace{1.1cm}
\includegraphics[scale=0.17, bb= 100 200 510 720, angle=90]{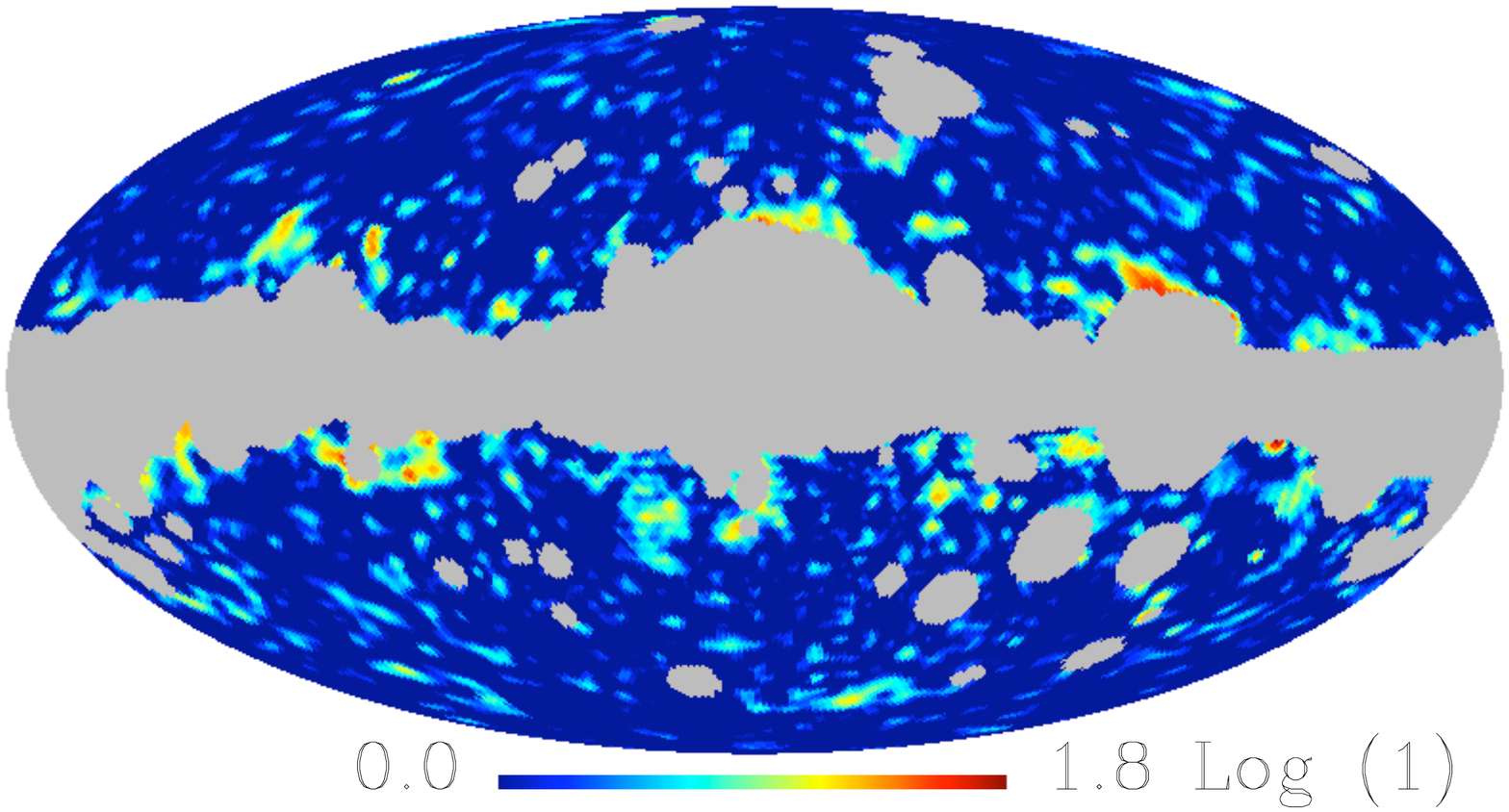}
\includegraphics[scale=0.17, bb= 100 200 510 700, angle=90]{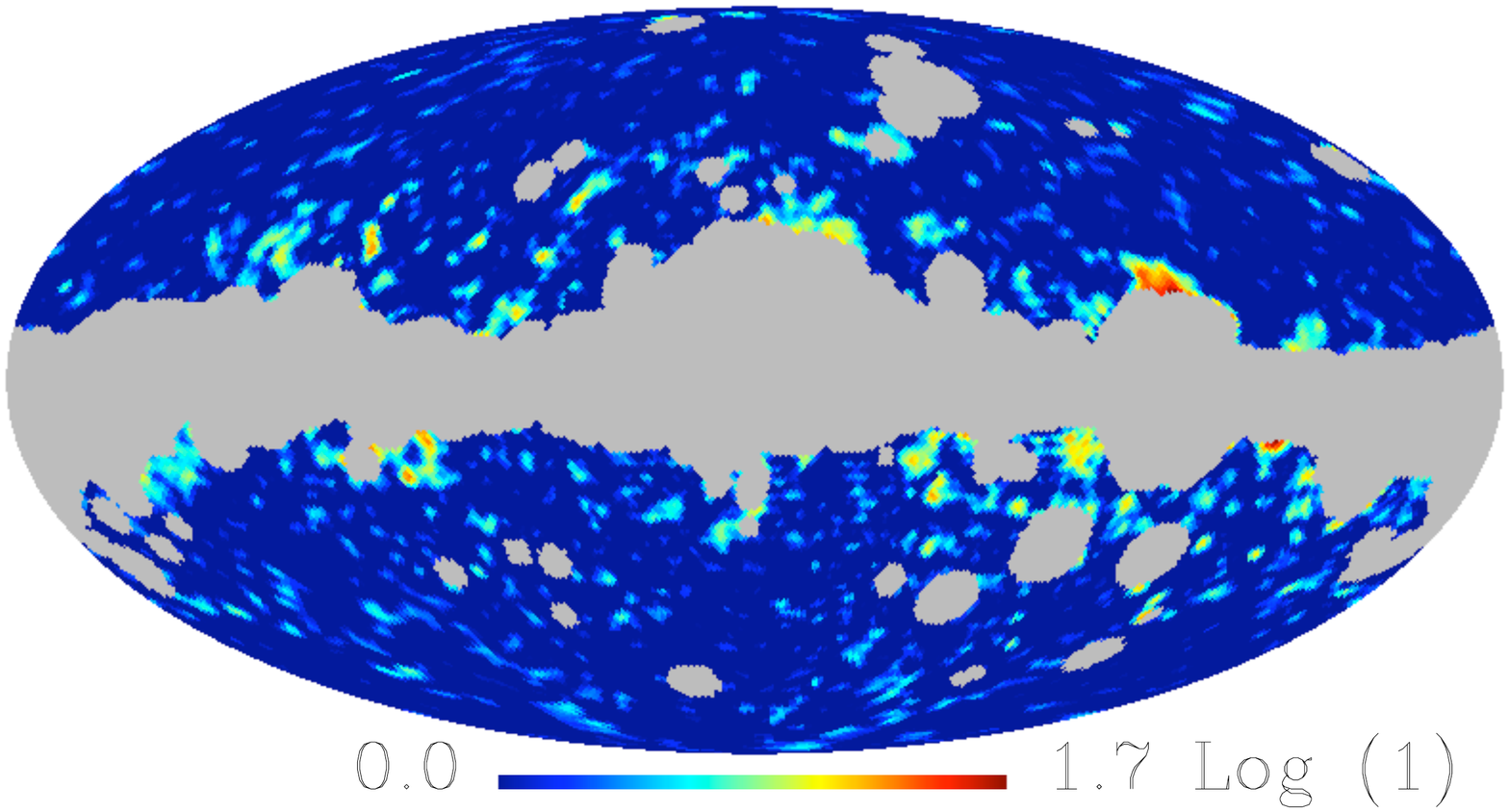}
\hspace{1.1cm}
\includegraphics[scale=0.17, bb= 100 200 510 730, angle=90]{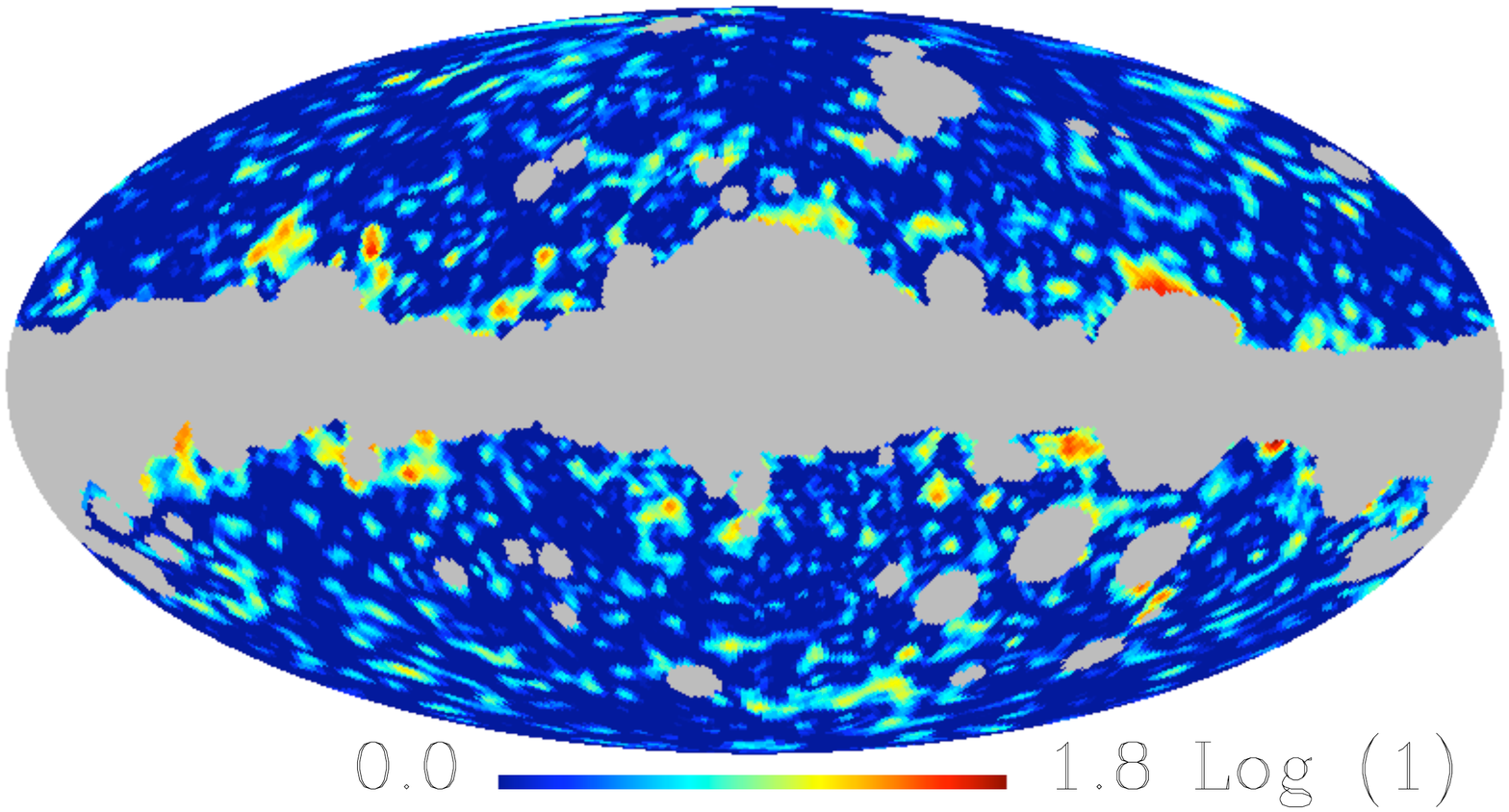}
\caption{ Maps of the residual emission (top 3 rows, presented in arbitrary units) and of the parameter $\zeta_{i,j}$ (bottom 3 rows) in the K (top), Ka (middle) and Q (bottom) frequency bands resulting from our 4-template fit when using the Gibbs (left) and ILC (right) CMB estimators, where for the maps of $\zeta_{i,j}^2$ we truncate the maps using a minimum of $\zeta_{i,j}^2=1$.}
\label{fig:diffzeta4temp:ch3}
\vspace{-0.5cm}
\end{center}
\end{figure}

In fig.~\ref{fig:chisqvstheta:ch3} we display the results for $\chi^2_{\rm red.}$ associated with 4-template fits for different values of the angle $\theta$. We find that the minimum value $\chi^2_{\rm red., min.}\simeq6.379~(9.173)$, occurs for $\theta\simeq48.5^{\circ} (46.5^{\circ})$ when using the Gibbs (ILC) CMB estimator. Thus, using our 4-template fit we have obtained a 16\% (17\%) reduction in the value of $\chi^2_{\rm red.}$ relative to that obtained with the 3-template fitting procedure discussed in \S\,\ref{sec:3temp}. 
\begin{figure}[h]
\begin{center}
\includegraphics[width=90mm, keepaspectratio, clip]{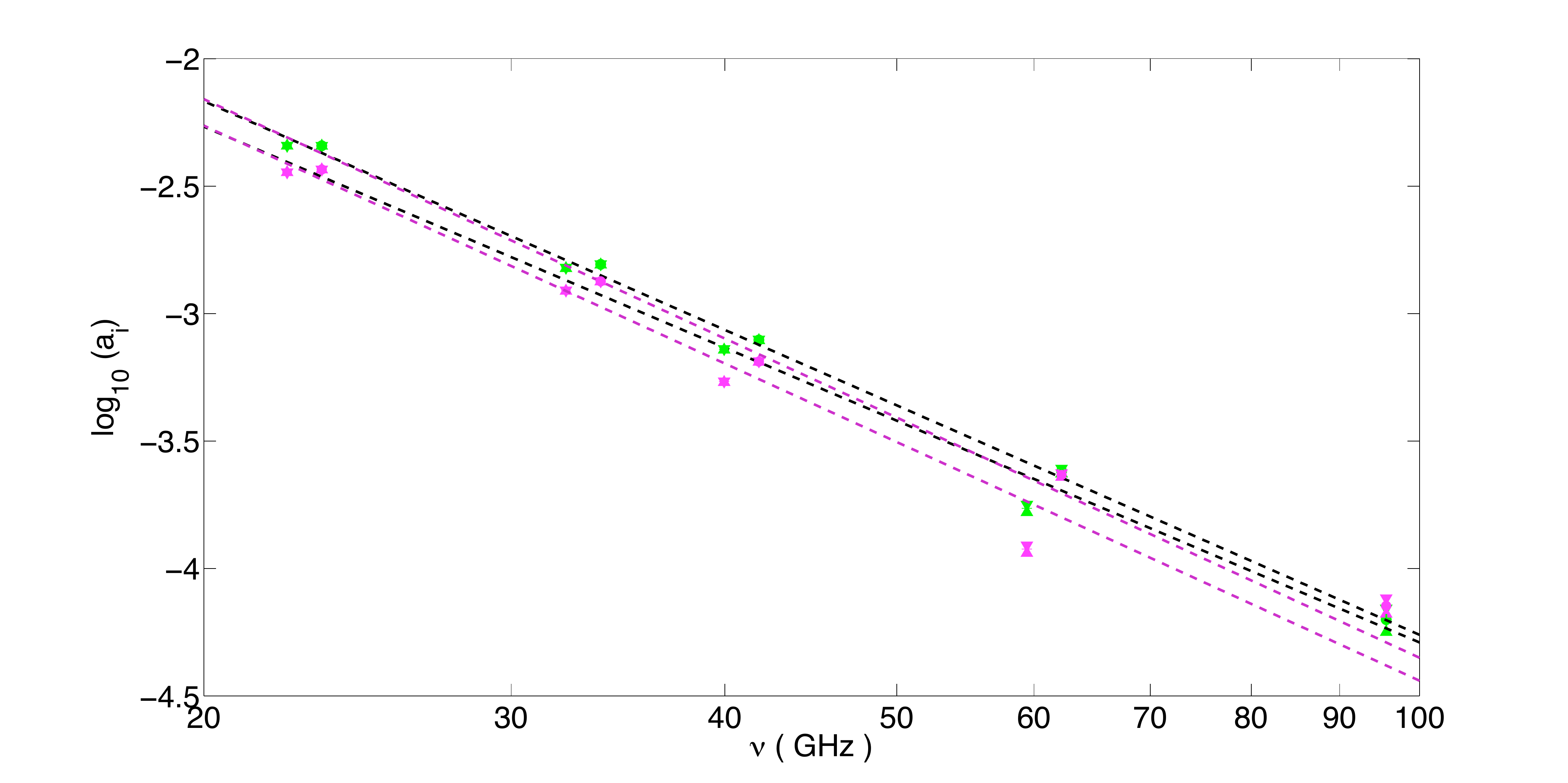}
\caption{Spectral dependences of best fitted soft synchrotron galactic foreground emission in the central (green markers) and peripheral (magenta markers) regions, resulting from our 4-template fit when using the Gibbs (circles) and ILC (crosses) CMB estimators. Also plotted are the lines of best fit associated with the best fit spectral indices for the displayed data sets (dashed black (Gibbs) and dashed magenta (ILC) ). For clarity, we have slightly displaced the frequency of the two sets of data markers associated the Gibbs and ILC CMB estimators by $\Delta{\rm log}_{10}(\nu /{\rm GHz})$=+0.5 and $-0.5$ respectively.}
\label{fig:spec4temp:ch3}
\end{center}
\end{figure}
\noindent This can be observed directly in fig.\,\ref{fig:diffzeta4temp:ch3}, where we display maps of the residual emission $\mathbf{r}$ and maps of the parameter $\zeta_{i,j}^2$ in the K (top), Ka (middle) and Q (bottom) frequency bands following the described 4-template fit when using the Gibbs (left) and ILC (right) CMB estimators, where again, for the maps of $\zeta_{i,j}^2$ we only display regions of statistically significant emission (i.e. $\zeta_{i,j}^2>1$). The values of the parameter $\zeta_i^2$ for the five full-sky residual maps are now 4.43 (5.36), 0.82 (1.35), 1.01 (2.02), 0.089 (0.17) and 0.028 (0.27) for the K, Ka, Q, V and W bands respectively. (The $1\sigma$ errors on these values are all of order $10^{-5}$.) These values correspond to decreases of 20.00\% (18.67\%), 7.70\% (6.77\%), 6.30\% (4.94\%), 0.05\% (1.40\%) and 0.06\% (0.35\%), relative to those obtained for the 3-template fit when using the Gibbs (ILC) CMB estimator. (The $1\sigma$ errors on these values are all of order $10^{-3}$ or less.) Again, we compare these results for $\zeta^2_{i}$ for the full-sky with those calculated using a region within $50^{\circ}$ of the GC, for which we obtain 7.94 (9.04), 1.24 (1.81), 1.20 (2.21), 0.107 (0.186) and 0.027 (0.29) for the K, Ka, Q, V and W bands respectively when using the Gibbs (ILC) CMB estimator. (The $1\sigma$ errors on these values are all of order $10^{-3}$ or less.) These values correspond to decreases in $\zeta_{i}^2$ of $45.98\pm0.12$\% ($45.52\pm0.11$\%), $24.82\pm0.15$\% ($24.89\pm0.12$\%), $25.24\pm0.16$\% ($22.00\pm0.10$\%), $0.23\pm0.01$\% ($7.81\pm0.08$\%) and $0.50\pm0.01$\% ($2.17\pm0.01$\%) in the K, Ka, Q, V and W bands respectively, relative to the corresponding 3-template fit results.

In fig.\,\ref{fig:spec4temp:ch3} we display the spectral dependences of the of the fitted central (green markers) and peripheral (magenta markers) synchrotron emission components when using the Gibbs (circles) and ILC (crosses) CMB estimators. Again, for clarity, we have slightly displaced the frequency of the two sets of data markers associated the Gibbs and ILC CMB estimators by $\Delta{\rm log}_{10}(\nu /{\rm GHz})$=+0.5 and $-0.5$ respectively. Since the differences in the spectral dependence of the two synchrotron components is very subtle, we have for clarity omitted plotting curves corresponding to the free-free and dust components, which for the current fit, were deduced to possess best fit spectral indices of -1.98 (-1.86) for free-free, -2.93 (-2.99) for dust (between K and Ka bands) and +1.14 (+1.23) for (between V and W bands), when using the Gibbs (ILC) CMB estimators. 

The respective best fit spectral indices for the central and peripheral synchrotron components were -2.99 (-3.13) and -2.89 (-3.11). (We have omitted the ILC results at 93.5 GHz which gave negative values for the synchrotron template weights, whose 1$\sigma$ errors were of the same order, further supporting the decision that such results should be deemed unreliable). Both of these values are consistent with the range determined in previous studies for the spectral index of soft synchrotron emission \cite{syncindex:ch3}. Further, both indices are extremely similar in themselves, contrary to the suggestion by Hooper, Finkbeiner and Dobler (2007) (HFD henceforth) that the haze emission possesses a significantly harder spectral dependence $I_{\nu}\propto\nu^{-0.25}$ \cite{HFD2007:ch3} than the conventional soft synchrotron emission far from the GC. 
Also, from fig.\,\ref{fig:spec4temp:ch3} it is clear that both regions of synchrotron emission are always very similar in intensity, and considering that their spectral indices are also very similar it is quite possible that our results simply indicate the slight regional variation in the conventional soft synchrotron emission (i.e. that associated with the 408\,MHz measurements by Haslam {\it et~al.}), rather than evidence for the presence of an independent, exotic source of synchrotron emission near the GC. Therefore, we have revealed the interesting result that a significant residual haze emission may be reproducible by simply invoking a slight regional dependence in the spectral index of the soft synchrotron foreground emission.

\section{Synchrotron radiation from dark matter annihilation}

One of the most heavily investigated extensions  to the standard model of particle physics is supersymmetry (SUSY), which is a framework based around a proposed symmetry between fermions and bosons. Moreover, in supersymmetric theories, for every boson (fermion) in the standard model there exists a corresponding fermionic (bosonic) superpartner that is typically much heavier than its standard model counterpart. Such theories have great appeal since they can potentially solve the so-called hierarchy and unification paradoxes renowned in particle physics. (For a brief review of SUSY see \cite{bertone05} \S\,3.2 and references therein).

For our purposes, the greatest appeal of  SUSY is that it provides an excellent CDM candidate, known as the neutralino $\chi$, with properties that lead to a relic density $\Omega_{\rm CDM}h^2\sim0.1$, consistent with current experimental evidence \cite{Dunkley:2008ie}. Perhaps most importantly, these  properties arise {\it independently} of the any such astrophysical constraints, but simply from requirements based on their ability to solve the hierarchy problem. 

The neutralino is a superposition of higgsinos, winos and binos, which are the fermionic superpartners of the standard model gauge bosons, and hence results in a diverse range of properties largely unconstrained by present observations. Consequently the neutralino is electrically neutral and colourless, only interacting weakly and gravitationally, and therefore very difficult to detect directly. Hence the neutralino is aptly categorised as a {\it weakly interacting massive particle} or WIMP. 
In R-parity conserving supersymmetric models the lightest neutralino, being the lightest SUSY 
particle (LSP) is stable \cite{weinberg}. Consequently, in a scenario where present-day CDM exists as a result of  thermal-freeze out, the dominant species of CDM is likely to include the LSP.  The relic density of the LSP will then heavily depend on its mass and self-annihilation cross-section.

In this section we calculate the expected flux in synchrotron radiation resulting from relativistic electrons/positrons $e^{\pm}$ produced by annihilating neutralino dark matter. In order to calculate such fluxes we need several ingredients. Firstly, in \S\,\ref{subsec:injectspec:ch3} we calculate the energy spectrum of $e^{\pm}$ injected into the ISM as a result of the annihilation of supersymmetric (SUSY) neutralinos with described by a specified parameter set. Secondly, in \S\,\ref{subsec:diffeq:ch3} we calculate the steady-state number density of $e^{\pm}$, for all relevant energies, at all positions along a particular line of site (l.o.s.), when accounting for the effects of galactic diffusion. Thirdly, in \S\,\ref{subsec:singlesyncP:ch3} we calculate the expected power in synchrotron radiation emitted by each electron at the frequency at which we are observing (i.e. one of the five WMAP frequencies). Lastly, in \S\,\ref{subsec:losflux:ch3} we use these results to calculate the total integrated power in synchrotron radiation along a given l.o.s. for a specified model of dark matter. We then compare our estimates for the synchrotron flux for several benchmark neutralino dark matter models with our estimates for the residual emission resulting from our 3-template fit presented in \S\,\ref{sec:3temp}.

\subsection{Positron injection spectra of annihilating SUSY neutralinos}
\label{subsec:injectspec:ch3}

We use four benchmark neutralino models that sample the vast particle parameter space of the minimal SUSY model (MSSM), owing to the fact that neutralinos are a superposition of four different supersymmetric particles (i.e. winos, binos and two higgsinos).
We use four benchmark models, the properties of which are as follows:
Firstly, we consider a 50\,GeV neutralino which annihilates 96\% of the time to $b{\bar b}$ pairs and 4\% to $\tau^+\tau^-$ (model~1). Secondly, we consider a 150\,GeV gaugino-dominated neutralino annihilating with identical branching ratios and modes as model\,1(model~2). Thirdly, we consider a 600\,GeV gaugino-dominated neutralinos that annihilate $87\%$ to $b{\bar b}$ and $13\%$ to $\tau^+\tau^-$ (model~3).
Lastly, we consider a 150\,GeV higgsino-dominated neutralinos, that annihilate $58\%$ to $W^+W^-$ and $42\%$ to $ZZ$  (model~4). 

\subsection{Calculation of steady-state positron distribution}
\label{subsec:diffeq:ch3}

The spectrum of $e^{\pm}$ resulting from the particle cascade immediately following the annihilation of two dark matter particles is subjected to significant modifications as these $e^{\pm}$ interact with the ISM. To calculate the spectrum of $e^{\pm}$ resulting from diffusion to any given point in the galaxy, we solve for the steady-state solution to diffusion-loss equation

\vspace{-0.5cm}
\be
\frac{\partial }{\partial t}\frac{{\rm d}n_e}{{\rm d}\epsilon}=\del\cdot\left[K(\epsilon, {\bf r})\del\frac{{\rm d}n_e}{{\rm d}\epsilon}\right]+\frac{\partial }{\partial \epsilon}\left[b(\epsilon, {\bf r})\frac{{\rm d}n_e}{{\rm d}\epsilon}\right]+Q(\epsilon, {\bf r}),
\label{eq:diffeq:ch3}
\ee
\vspace{0.1cm}

\noindent where ${{\rm d}n_e}/{{\rm d}E_e}$ is the number density of $e^{\pm}$ per unit energy, $K(E_e, {\bf r})$ is the spatial diffusion parameter, $b(E_e, {\bf r})$ is the rate of energy loss and $Q(E_e, {\bf r})$ is the source term which is proportional to the local rate of dark matter annihilation. We solve (\ref{eq:diffeq:ch3}) using exactly the same procedure as that described in \cite{baltz_edsjo}. Therefore, for brevity, we omit a discussion of the full procedure here, instead simply describing the parameter sets we utilised in our analysis. 

For diffusion and energy loss parameters of the form
\be
K(\epsilon, {\bf r})=K_0\epsilon^{\alpha}\quad~b(\epsilon, {\bf r})=\frac{\epsilon^2}{\tau_E},
\ee
the general form of our solutions for the steady-state number density of $e^{\pm}$, for a dark matter particle mass $m_{\chi}$, thermally-averaged annihilation cross-section (multiplied by velocity) $\langle\sigma_{\rm ann.}v\rangle$ and $e^{\pm}$ injection spectrum (per annihilation) d$\phi/$d$\epsilon$, given by
\bea
\frac{{\rm d}n_e}{{\rm d}\epsilon}&=&\frac{1}{2}\left(\frac{\rho_s}{m_{\chi}}\right)^2\langle\sigma_{\rm ann.}v\rangle\tau_E\epsilon^{-2}\int\limits^{m_{\chi}}_{\epsilon}{\rm d}\epsilon' D^{-2}\frac{{\rm d}\phi}{{\rm d}\epsilon'}\nonumber\\
&\times&\int\limits^{\infty}_{0}{\rm d}r'r'f(r')
{\rm exp}\left(-\frac{\left(r^2+r'^2\right)}{D^2}\right)I_0\left(\frac{2rr'}{D^2}\right)\nonumber\\
&\times&\sum\limits^{\infty}_{n=-\infty}(-1)^nF(\epsilon, z)\nonumber\\
\label{eq:dndE:ch3}
\eea
where
\bea
\nu&=&\frac{1}{\left(1-\alpha\right)}\epsilon^{(\alpha-1)},\nonumber\\
\epsilon&=&\frac{E_{e}}{1\,{\rm GeV}},\nonumber\\
D&=&\left[4K_0\tau_E\left(\nu(\epsilon)-\nu(\epsilon')\right)\right]^{1/2}\quad{\rm and}\nonumber\\
F(\epsilon, z)&=&{\rm erf}\left(\frac{L+(-1)^n(2Ln-z)}{D}\right)\nonumber\\
&&-{\rm erf}\left(\frac{-L+(-1)^n(2Ln-z)}{D}\right).\nonumber\\
\eea
The function $f(r)$, where $r$ is the cylindrical radial coordinate, is equal to our choice of spherical density profile, divided by the scale density $\rho_s$, averaged over the full azimuthal dimension of the diffusion zone (conventionally defined to be the region $r>0, |z|\le L$ containing a homogeneous and isotropic magnetic field)
\bea
f(r)&=&\frac{1}{2L}\int\limits^L_{-L}{\rm d}z\left(\frac{(r^2+z^2)^{1/2}}{r_s}\right)^{-2\gamma}\nonumber\\
&\times&\left[1+\left(\frac{(r^2+z^2)^{1/2}}{r_s}\right)^{\alpha}\right]^{\frac{-2(\beta-\gamma)}{\alpha}}.\nonumber\\
\eea

As an example, we illustrate the characteristics of the steady-state solution by using the simplistic set of parameters similar to that utilised by Finkbeiner (i.e. model~1 of \cite{Finkbeiner2005:ch3}) \cite{footnote7}
:$~K=K_0=9\times10^{27}$cm$^2$s$^{-1}$, $b=\tau_E^{-1}=10^{-16}$s$^{-1}$. We use a spherical NFW dark matter density profile \cite{nfw1996,nfw1997}, i.e. $(\alpha, \beta, \gamma)=(1,3,1)$, with a scale radius and density $r_s=20\,$kpc and $\rho_s=5.6\times10^{-25}\,$g~cm$^{-3}$ respectively. Finally, we use a 50\,GeV dark matter particle with annihilation cross-section given by $\langle\sigma_{\rm ann.}v\rangle=2\times10^{-26}$cm$^3$s$^{-1}$ which self-annihilates entirely to $e^+e^-$. This value of $\langle\sigma_{\rm ann.}v\rangle$ is very similar to the canonical value of $3\times10^{-26}$\,cm$^3$\,s$^{-1}$ consistent with dark matter relic density estimates ($\Omega_{\rm CDM}h^2=0.1099\pm0.0062$) from WMAP \cite{Dunkley:2008ie}.
\begin{figure}
\begin{center}
\includegraphics[width=\linewidth, keepaspectratio, clip]{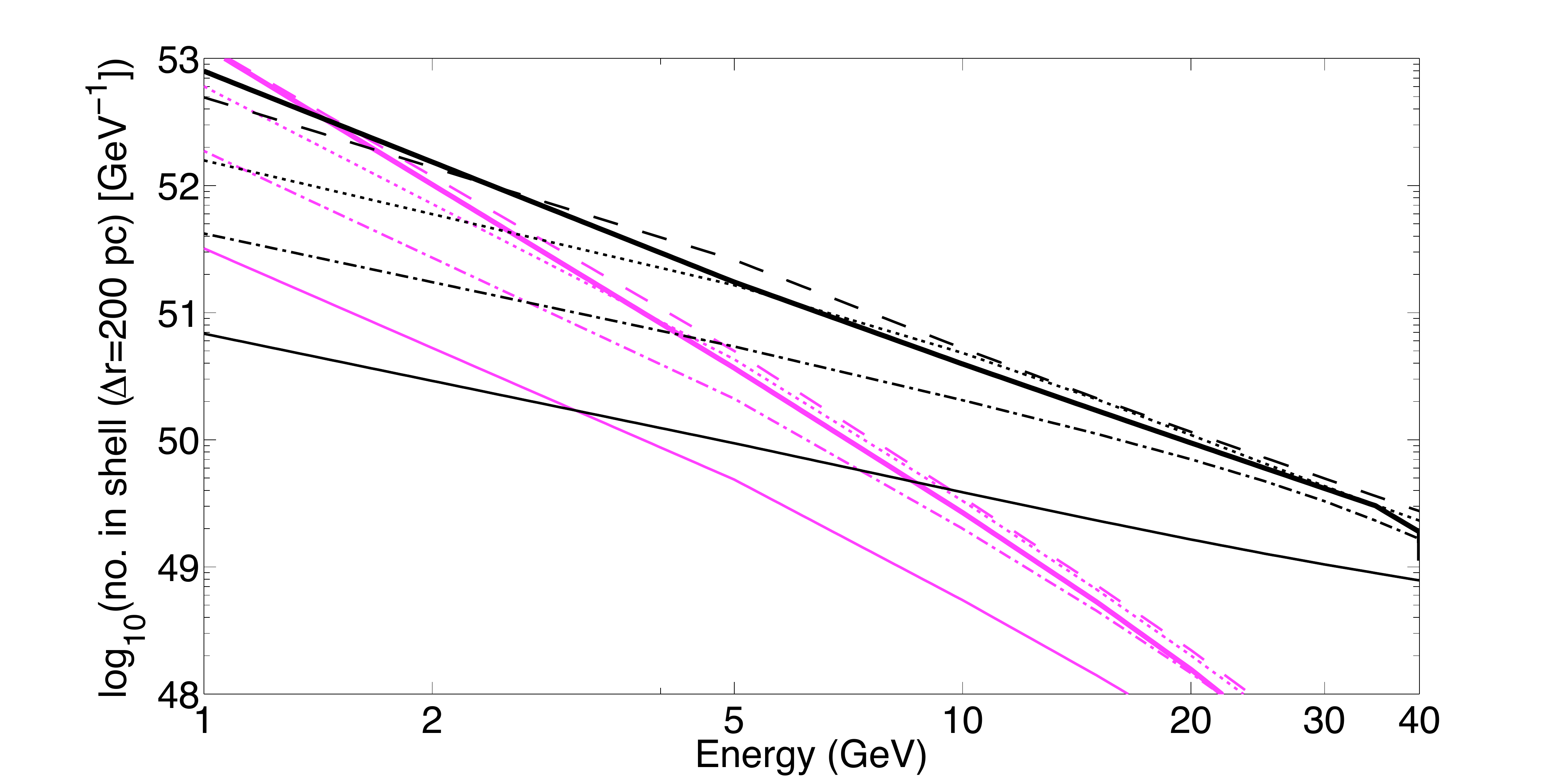}
\caption{Positron energy distribution for all positrons contained within spherical shells of thickness $\Delta r=200~$pc located at radii 8 (thick solid), 4 (dashed), 2 (dotted), 1 (dot-dashed) and 0.4 (thick solid) kiloparsecs from the Galactic centre, calculated using (i) 50\,GeV neutralinos with a monochromatic positron spectrum at  50\,GeV (black curves) (ii) 50\,GeV neutralinos with branching ratios $B(\chi\chi\rightarrow b{\bar b})=0.96$ and $B(\chi\chi\rightarrow\tau^+\tau^-)=0.04$ (magenta curves).}
\label{fig:noinshell:ch3}
\end{center}
\end{figure}
In fig.\,\ref{fig:noinshell:ch3} we display the resulting $e^{\pm}$ number density as a function of energy, integrated over shells of thickness $\Delta r=200\,$pc, plotted at radii $r=$8 (thick solid), 4 (dashed), 2 (dotted), 1 (dot-dashed) and 0.4 (thick solid) kiloparsecs (black curves). These results are clearly consistent with those presented in \cite{Finkbeiner2005:ch3}.
In fig.\,\ref{fig:noinshell:ch3} we also display the corresponding results for our benchmark model~1, that is, a 50\,GeV Gaugino-dominated neutralino with branching modes $B(\chi\chi\rightarrow b{\bar b})=0.96$ and $B(\chi\chi\rightarrow\tau^+\tau^-)=0.04$ and annihilation cross-section $\langle\sigma_{\rm ann.}v\rangle=2.7\times10^{-26}$cm$^3$s$^{-1}$ (magenta curves). 

Comparing the two sets of results we observe that the injection spectrum of positrons resulting from dark matter annihilation plays a significant role in the corresponding steady-state distribution throughout the Galaxy. As we expect, the more realistic steeply declining (with respect to energy), injection spectrum from our model\,1 neutralinos results in a positron distribution which also steeply declines with energy at a rate much faster than that associated with neutralinos with a monochromatic injection spectrum. Such observations prove extremely significant when we consider the synchrotron radiation flux produced by these particles since in the following subsection we will demonstrate that only $e^{\pm}$ {\it above a certain energy threshold} effectively contribute to the synchrotron flux at any given frequency.

\subsection{Synchrotron radiation spectrum}
\label{subsec:singlesyncP:ch3}

Synchrotron radiation from $e^{\pm}$ results from their acceleration by the Galactic magnetic field that defines the diffusion zone. For $e^{\pm}$ possessing a Lorentz factor $\gamma=E/m_e$, confined to helical trajectories with an isotropical distribution of pitch angles $\alpha$, with respect to a local magnetic flux $B$, the average fractional power in synchrotron radiation emitted at frequency $\nu$ is given by \cite{Rybicki_Lightman:ch3, Baltz2004:ch3}:
\be
\frac{1}{P_{\rm sync.}^{\rm tot.}}\frac{{\rm d}p_{\rm sync}(\nu)}{{\rm d}x}=\frac{27\sqrt{3}}{32\pi}x\int\limits^{\pi}_{0}{\rm d}\alpha\int\limits^{\infty}_{x/{\rm sin}\alpha}{\rm d}y K_{5/3}(y),
\label{eq:fracP:ch3}
\ee
where $x=\nu/\left(\nu_B\gamma^2\right)$, with $\nu_B=3eB/(4\pi m_ec)\simeq42\left(B/10\,\mu{\rm G}\right)$Hz, and the total synchrotron power $P_{\rm sync.}^{\rm tot.}$ is set equal to a fraction $f_{\rm sync.}$ of the total emitted
power $b(\epsilon)$
\bea
P_{\rm sync.}^{\rm tot.}&=&f_{\rm sync.}b(\epsilon)=f_{\rm sync.}\frac{\epsilon^2}{\tau_E}\nonumber\\
&\equiv&\frac{\epsilon^2}{\tau_{\rm sync.}},
\eea
where $\tau_{\rm sync.}=\tau_E/f_{\rm sync.}$ is the synchrotron energy loss time scale.

\begin{figure}
\begin{center}
\includegraphics[width=\linewidth, keepaspectratio, clip]{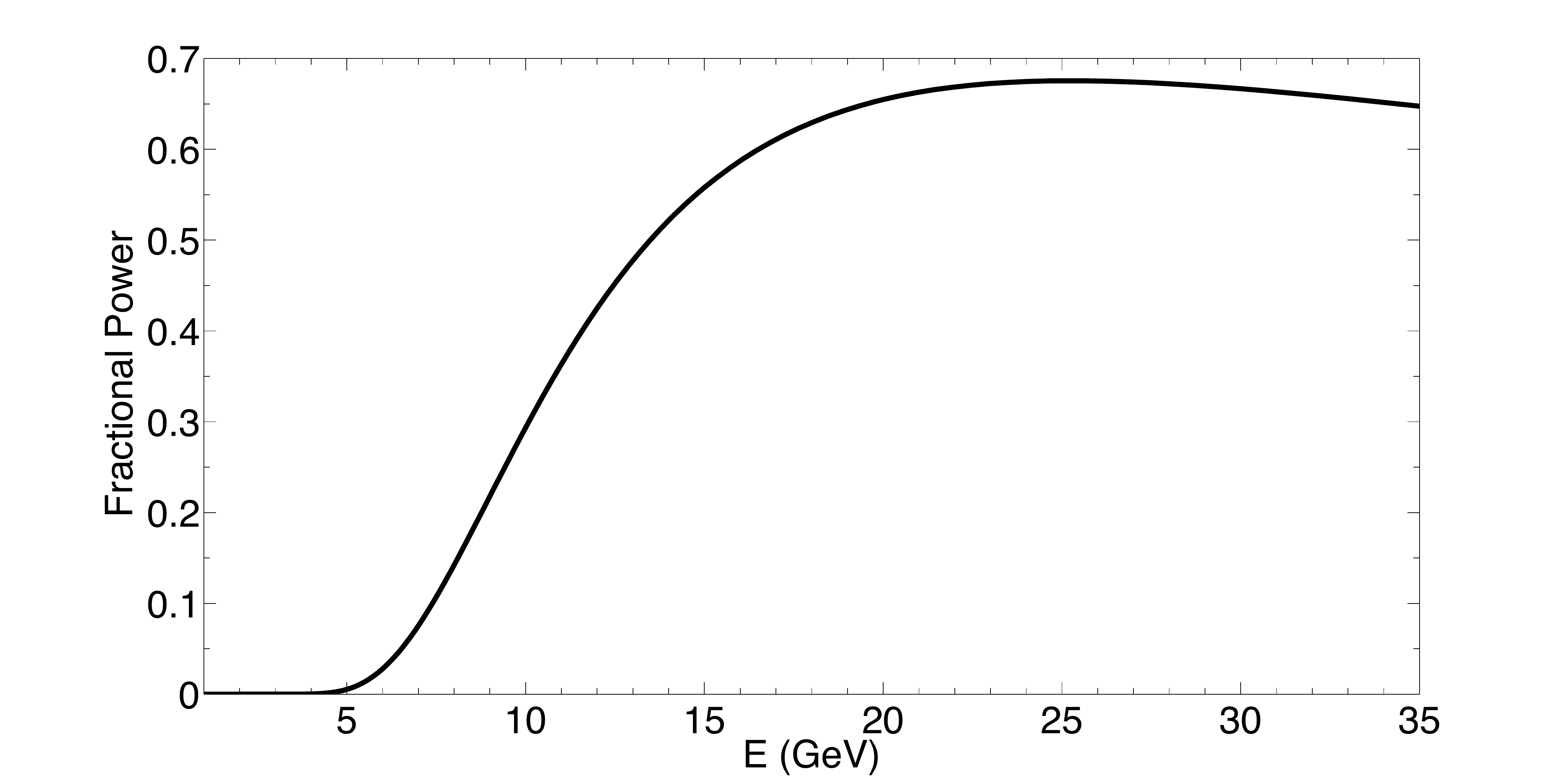}
\includegraphics[width=\linewidth, keepaspectratio, clip]{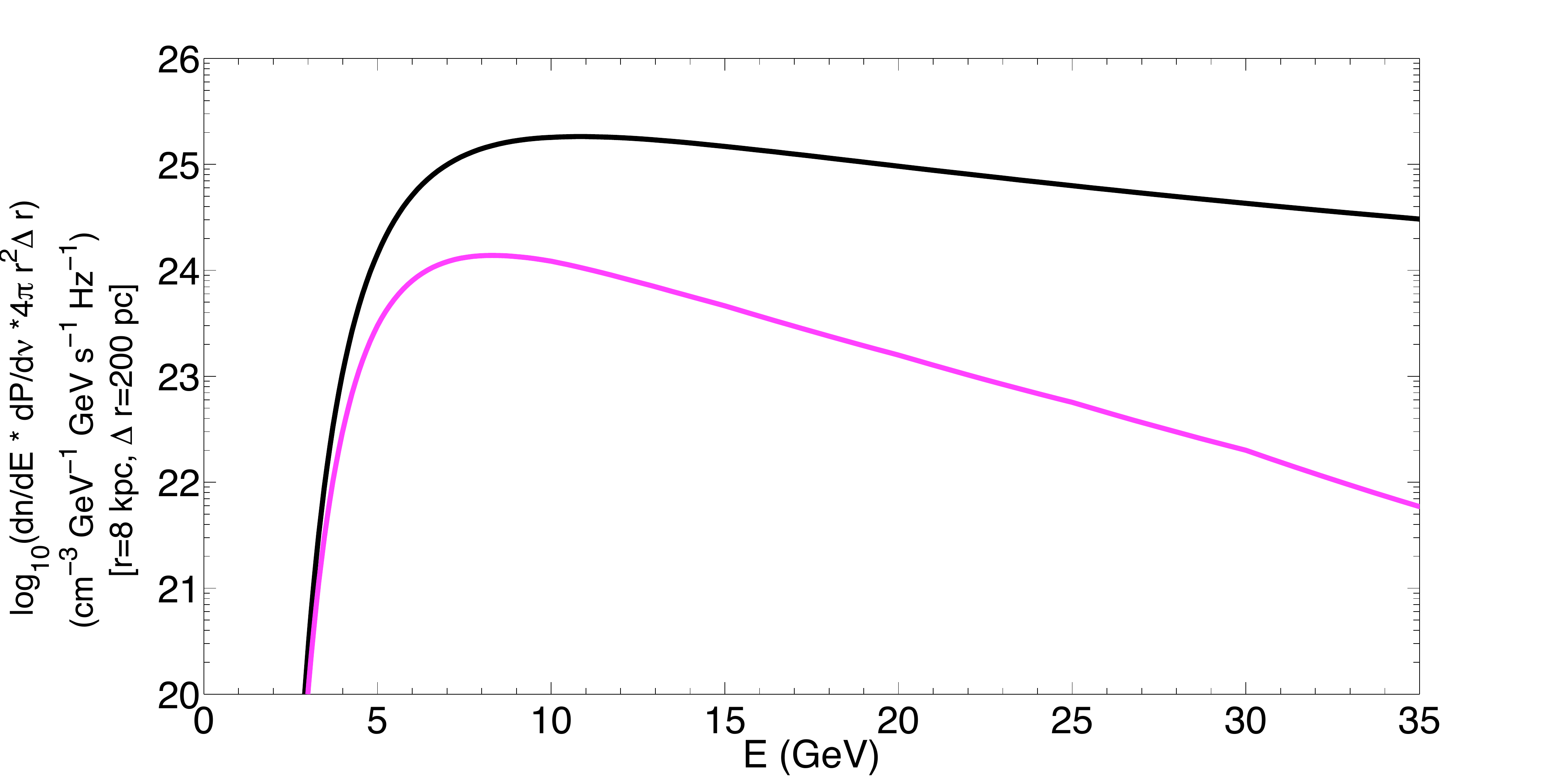}
\caption{The expected synchrotron power to eminate from a spherical shell of thickness 200\,pc and inner radius 8\,kpc, at the K-band frequency of 22.8\,GHz for our realistic neutralino model (magenta curve) less-realistic model involving direct annihilation channels to electron/positrons (black curve).   }
\label{fig:fracsyncP:ch3}
\end{center}
\end{figure}

In fig.~\ref{fig:fracsyncP:ch3} we plot the spectrum in fractional power (\ref{eq:fracP:ch3}) radiated by electrons in the presence of a magnetic field of strength $B=10\,\mu{\rm G}$ at frequency 22.8\,GHz (K-band).  Only electrons with energies above a certain threshold $\gamma^2>\nu/\nu_B\simeq12\,$GeV (i.e. $x<1$), contribute efficiently to the radiated power. 
To illustrate the significance of this in our present analysis, in fig.\,\ref{fig:fracsyncP:ch3} (bottom panel) we display the emissivity (i.e. number density multiplied by emitted power) in synchrotron radiation at 22.8\,GHz for a Galactocentric radius of 8\,kpc, for the monochromatic 50\,GeV neutralino dark matter (black line) and the 50\,GeV neutralino dark matter described by our model~1 (magenta line). Despite the fact that the model~1 neutralinos generate a proportionally larger number of low energy $e^{\pm}$, which effectively do not contribute to the synchrotron power, its steeper injection spectrum results in a significantly smaller contribution (by at least an order of magnitude) from higher energy $e^{\pm}$ compared to the monochromatic $e^{\pm}$ produced by our alternative neutralino model.

\subsection{Results for synchrotron flux}
\label{subsec:losflux:ch3}

We now have all the necessary ingredients in order to calculate the synchrotron flux resulting from annihilation of Galactic dark matter along a given l.o.s.. To obtain such a flux, we firstly calculate the column density spectrum of $e^{\pm}$, ${\rm d}\sigma/{\rm d}\epsilon$ by integrating (\ref{eq:dndE:ch3}) along a specified l.o.s. 
\be
\frac{{\rm d}\sigma(\delta, \epsilon)}{{\rm d}\epsilon}=\int\limits^{l_{\rm max.}}_{l=0}{\rm d}l\frac{{\rm d}n_e}{{\rm d}\epsilon},
\label{eq:dsigmadE:ch3}
\ee
\noindent where $l_{\rm max.}=L/|{\rm sin}(\delta)|$ is the path length along our l.o.s., which lies within a plane intersecting the GC and with inclination $\delta$ relative to the GC. The spectrum in synchrotron power ${\rm d}P_{\rm sync.}/{\rm d}\nu$ from the full column of $e^{\pm}$ along a given l.o.s. is then simply (in power per area per unit solid angle per frequency interval) obtained by integrating (over energy) the column depth (\ref{eq:dsigmadE:ch3}) with the synchrotron power spectrum for a single electron, ${\rm d}p_{\rm sync.}/{\rm d}\nu$, given by (\ref{eq:fracP:ch3}),
\bea
\frac{{\rm d}P_{\rm sync.}(\delta, \epsilon)}{{\rm d}\nu}&=&\frac{1}{4\pi}\int\limits_{0}^{m_{\chi}}{\rm d}\epsilon~\frac{{\rm d}p_{\rm sync.}(\nu, \epsilon)}{{\rm d}x}\frac{{\rm d}x}{{\rm d}\nu}~\frac{{\rm d}\sigma(\delta, \epsilon)}{{\rm d}\epsilon}\nonumber\\
&=&\frac{27\sqrt{3}}{128\pi^2}\frac{m_e^2}{\tau_{\rm sync.}\nu_B}\int\limits^{m_{\chi}}_{0}{\rm d}\epsilon~\epsilon^2\frac{{\rm d}\sigma(\delta, \epsilon)}{{\rm d}\epsilon}\frac{x}{\nu}\nonumber\\
&\times&~x\int\limits^{\pi}_{0}{\rm d}\alpha\int\limits^{\infty}_{x/{\rm sin}\alpha}{\rm d}y K_{5/3}(y).\nonumber\\
\eea
So that we can compare our results with those obtained using the treatment by Hooper, Finkbeiner \& Dobler \cite{HFD2007:ch3}, we use the following set of astrophysical parameters: $K_0=10^{28}$cm$^3$s$^{-1}$, $\alpha=0.33$, $\tau_E=2\times10^{15}$s, $f_{\rm sync.}=0.25$, $B=10\,\mu$G, $L=3\,$kpc.
\begin{figure}
\begin{center}
\includegraphics[width=90mm, keepaspectratio, clip]{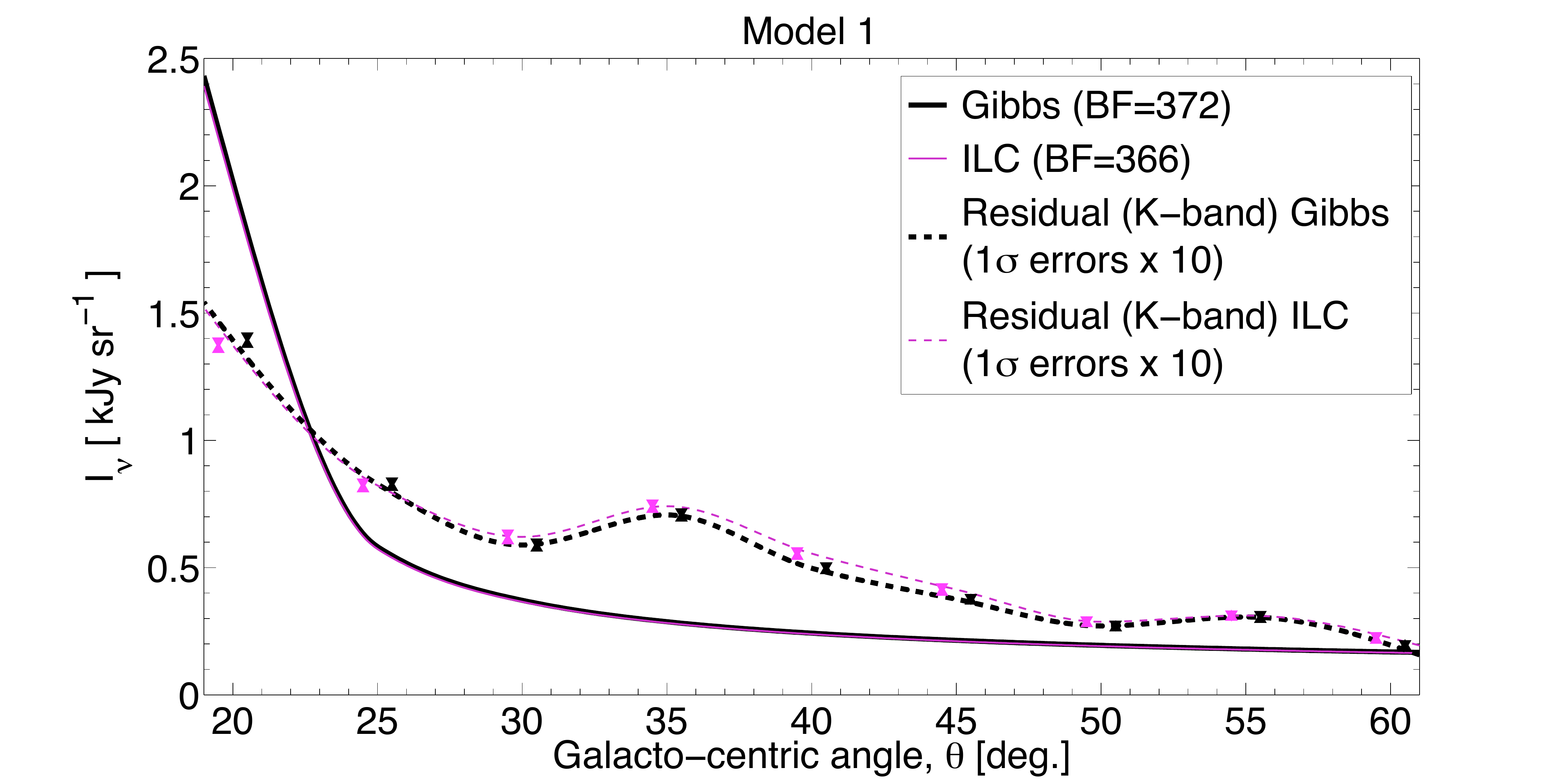}
\includegraphics[width=90mm, keepaspectratio, clip]{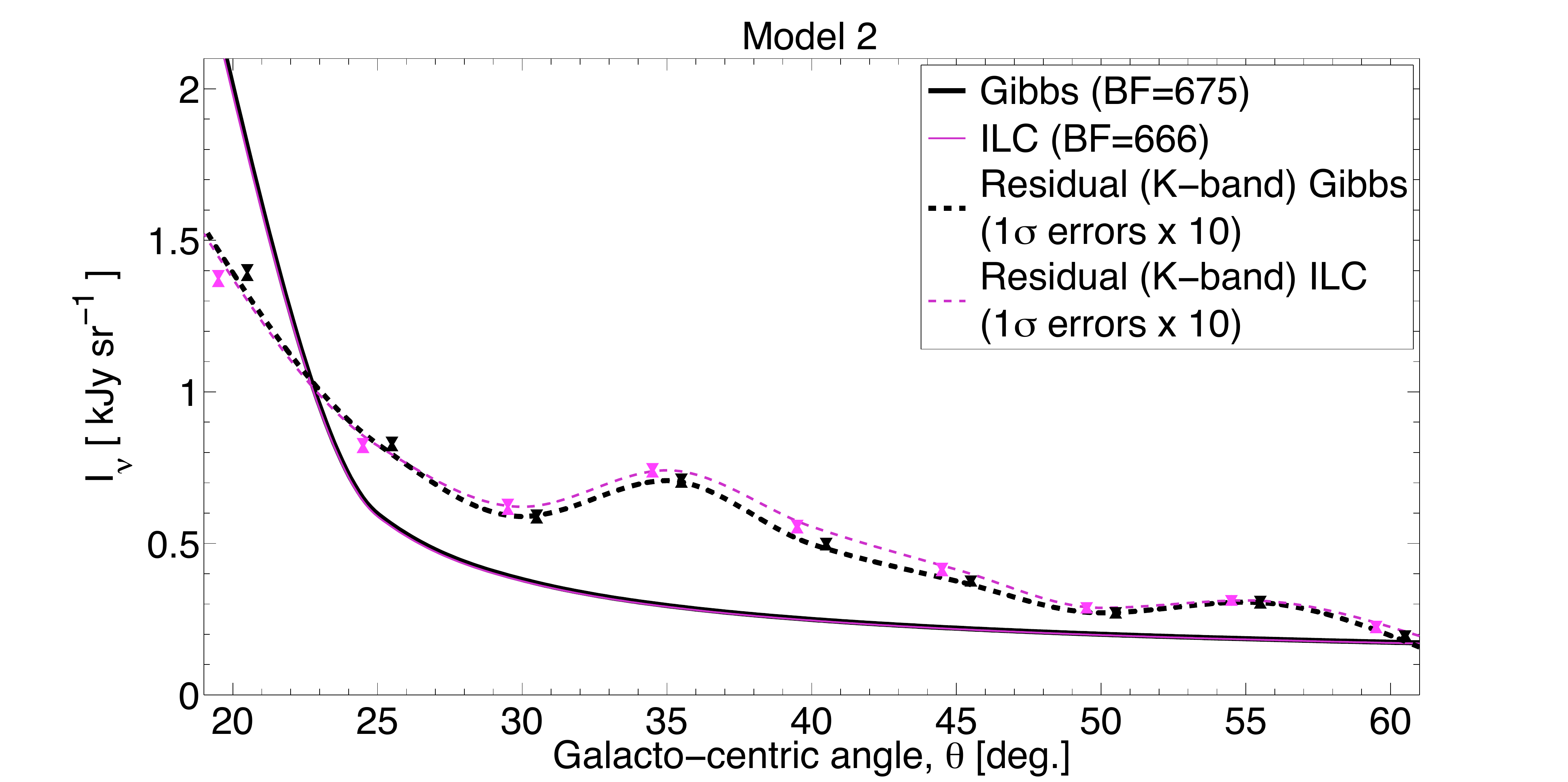}
\includegraphics[width=90mm, keepaspectratio, clip]{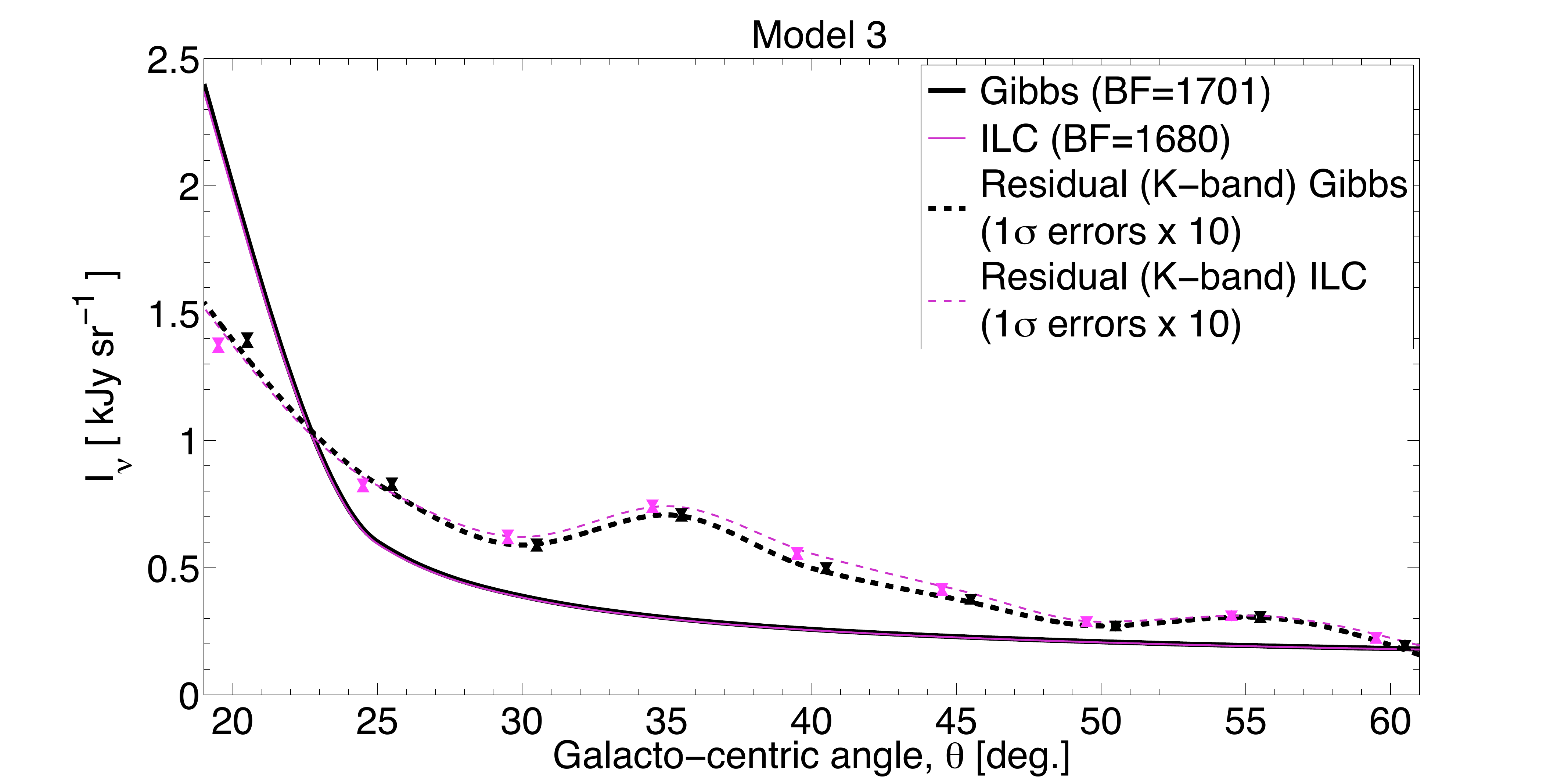}
\includegraphics[width=90mm, keepaspectratio, clip]{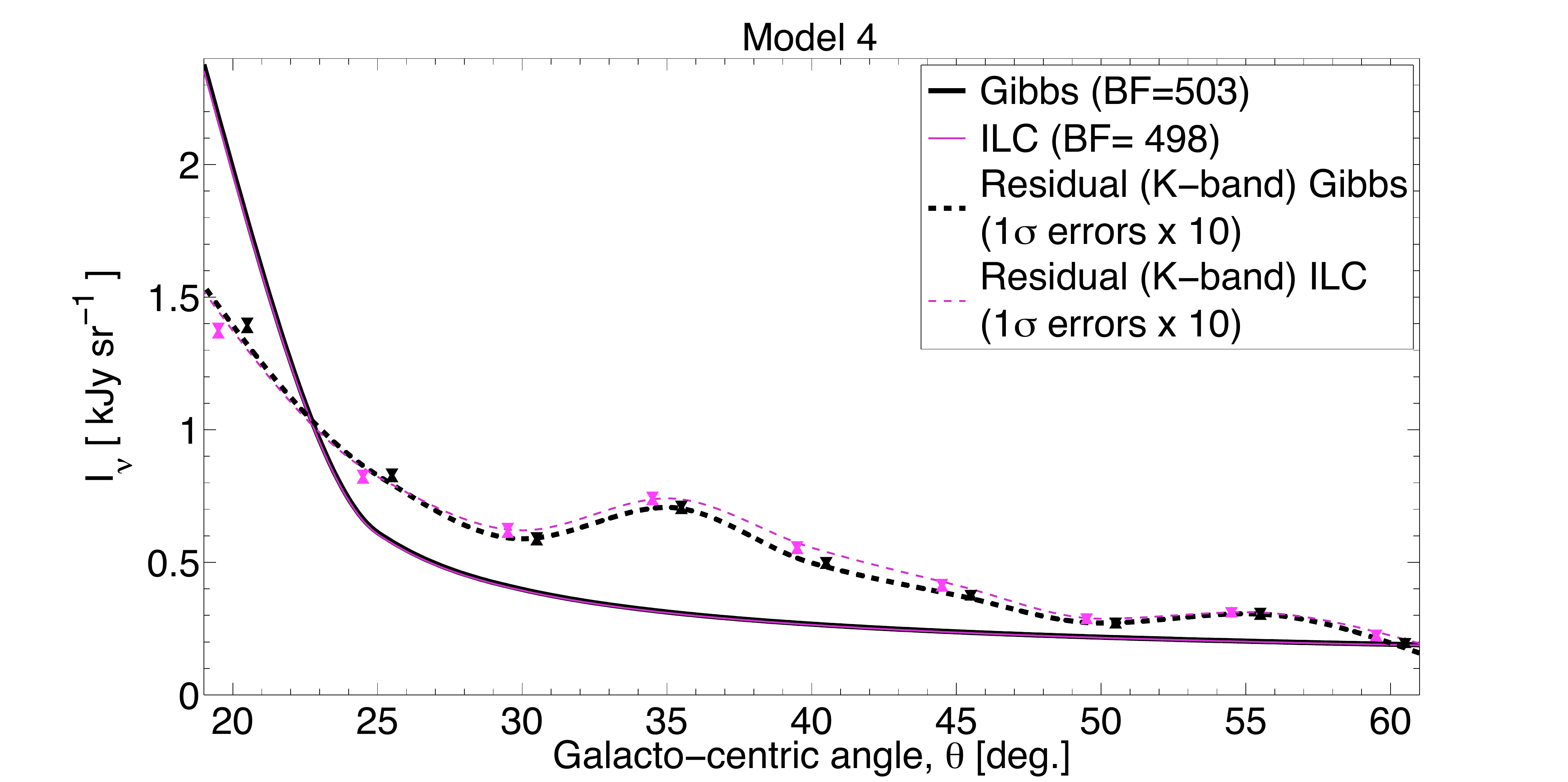}
\caption{Estimated synchrotron emission from dark matter annihilation using our four benchmark models  1-4 (top to bottom respectively), re-normalised, via a boost factor BF, to the deduced residual emission in the K-band.}
\label{fig:DM_flux:ch3}
\end{center}
\end{figure}
In fig.\,\ref{fig:DM_flux:ch3} we compare our estimates for the angular distribution of synchrotron power in the K-band resulting from the annihilation of dark matter (solid curves), for all four benchmark models, with that obtained from our 3 template-based fit to the CMB-subtracted WMAP data when using the Gibbs (black markers) and ILC (magenta markers) CMB estimators, where once again we have slightly displaced the frequency of the two sets of data markers. We have also plotted the lines of best fit to the deduced K-band residual emission (dashed curves). The noise in the measurements of the sky temperature is very small at the resolution that we use here.  In consequence the $1\sigma$ statistical error  
bars on CMB measurements associated with the displayed residual emission are also  very small. Therefore, in order to illustrate these errors we have displayed the 1$\sigma$ error bars, deduced in the procedure described above, enlarged by a factor of 10. We have multiplied each dark matter flux by a universal normalising factor, or {\it boost factor} $BF$, in order to obtain the best-fit to the WMAP data, where for each model we minimise the parameter
\vspace{0.3cm}
\be
\chi^2=\frac{\left[BF\times\left(\frac{{\rm d}P_{\rm sync.}(\theta)}{{\rm d}\nu}\right)_{\rm DM}-\left(\frac{{\rm d}{\bar P_{\rm sync.}}(\theta)}{{\rm d}\nu}\right)_{\rm haze}\right]^2}{{\bar \sigma^2}(\theta)},
\ee
\vspace{0.3cm}

\noindent where $\left({\rm d}{\bar P}_{\rm sync.}(\theta)/{\rm d}\nu\right)_{\rm haze}$ and ${\bar \sigma^2}(\theta)$ are the differential residual synchrotron power and (1$\sigma$) error map (squared) in the CMB-subtracted data in the K-band, averaged over the beam FWHM at Galacto-centric inclination $\theta$. 

The flux produced from annihilations within a smooth distribution of Galactic dark matter is far less than that required to explain the WMAP haze. This is contrary to the results presented by HFD as well as Caceres \& Hooper \cite{Caceres08}, who claim that such annihilations provide a flux, which is of the order of several kJy at the relevant $\theta$ investigated, and therefore fits extremely well to the residual haze emission. Here, we find that we require boost factors of approximately 372 (366), 675 (666), 1701 (1680) and 503 (498) for models 1-4 respectively, in order to generate the necessary synchrotron power and explain our discerned haze emission in the K-band. 

Despite this, the angular distribution of the estimated emission from dark matter fits very well to the data (despite the cylindrical nature of our selected dark matter density profiles). Interestingly we see that the angular dependence of the emission is very similar for all four neutralino models, 
with values of $\chi^2$ equal to 94.4\% (98.5\%), 88.7\% (92.9\%) and 82.1\% (86.4\%) of that associated with model 1, for models 2, 3 and 4 respectively.   
This implies that the gaugino or higgsino fractions are quite irrelevant in the determination of the observed synchrotron power spectrum. Moreover, the values of the best-fit boost factors obtained support the notion that it is the mass of the neutralino which dictates the overall normalisation of the spectrum (i.e. we find smaller boost factors for lighter neutralinos, which possess a higher rate of annihilation since the annihilation rate scales as $m_{\chi}^{-2}$), whilst the dark matter profile of the Galaxy dictates the shape of the spectrum.

\section{Conclusions}
We have investigated the prospects for the existence of the WMAP haze and the potential for it to be explained with either a regional dependence in soft synchrotron foreground emission or synchrotron emission resulting from the diffusion of electrons and positrons produced by annihilating neutralino dark matter, and found that the former is a better explanation.
We conducted parallel analyses using both Gibbs and ILC CMB estimators and found that generally all our results presented are mostly insensitive to the choice of the Gibbs or ILC CMB estimator, but we  believe that the Gibbs estimator is somewhat more trustworthy than the ILC.

Using only a simple 4-template fit, with the synchrotron template cut into two independent regions with different spectral indices, we obtained a significant reduction in residual K-band emission compared to our initial 3-template fit.  The reduction was approximately 20\% (19\%) for our full sky analysis and 46\% (46\%) in the inner 50$^{\circ}$ surrounding the galactic centre, when using the Gibbs (ILC) CMB estimator. We also found that the spectral dependences of the central and peripheral soft synchrotron emission possessed very similar spectral dependences in order to produce this improved fit, implying that a small regional variation in the spectral dependence of the soft synchrotron foreground emission can account for a substantial proportion of the observed residual emission.

Using four benchmark models of SUSY neutralinos we calculated the flux in synchrotron radiation in the K-band (where the haze possesses the greatest statistical significance) resulting from  energy losses by electrons and positrons produced in dark matter annihilations as they diffuse through the ISM. We fitted the angular dependence of the l.o.s. flux with that of the residual emission determined from our 3-template fit.  We deduced that in all four models, significant enhancements in the annihilation rate were required in order to account for the observed residual emission. Specifically, the calculated synchrotron flux resulting from dark matter annihilations needed to be multiplied by a {\it boost factor} ranging from approximately 300-1700 in order to best-fit to the residual emission. 

The possibility that such boosting might be achievable when acknowledging the presence of dark matter substructures, which have been postulated to exist to various degrees by many recent  N-body simulations of dark matter halos \cite{n_body_sims}. This has been the subject of recent investigation by Borriello, Cuoco and Miele (BCM) \cite{bcm}. BCM conclude that even with such boosting conservative constraints on the annihilation cross-section are $\langle\sigma_{\rm ann.}v\rangle\sim10^{-23}$\,cm$^3$\,s$^{-1}$$-10^{-25}$\,cm$^3$\,s$^{-1}$, which is up to two orders of magnitude larger than the canonical value of $3\times10^{-26}$\,cm$^3$\,s$^{-1}$ consistent with dark matter relic density estimates from WMAP.

\section{acknowledgments} 
\noindent We would like to thank Jo Dunkley and Massimiliano Lattanzi for their helpful comments. We acknowledge use of the Legacy Archive for Microwave Background Data Analysis (LAMBDA) (http://lambda.gsfc.nasa.gov/). We acknowledge use of the HEALPix software (http://HEALPix.jpl.nasa.gov/) and analysis package for deriving the results in this paper. DTC is supported by CERCA Case Western Reserve University and funded by NASA grant 4200188792. DTC's research is also supported in part by the US DoE. JZ is supported by an STFC rolling grant. HKE acknowledges financial support from the Research Council of Norway.


\begin{thebibliography}{99}
\bibitem{footnote2}
Since we work at low resolution statistical noise in the data is minimal,  the 3 year WMAP data is therefore as good as the 5 year data for our purposes.
\end{thebibliography}
\end{document}